%% file: ms_v8.tex
\documentclass[]{emulateapj}

\bibpunct{(}{)}{;}{a}{}{,}

\slugcomment{ApJ, accepted}
\shorttitle{Segue 1: An Unevolved Fossil Galaxy from the Early Universe}
\shortauthors{Frebel, Simon \& Kirby}

\begin{document}

\title{Segue 1: An Unevolved Fossil Galaxy from the Early
  Universe\altaffilmark{*}}

\author{
Anna Frebel\altaffilmark{1},
Joshua D. Simon\altaffilmark{2},
and Evan N. Kirby\altaffilmark{3,4}}

\altaffiltext{*}{This paper includes data gathered with the 6.5 meter
  Magellan Telescopes located at Las Campanas Observatory, Chile.
  Data herein were also obtained at the W. M. Keck Observatory, which is
  operated as a scientific partnership among the California Institute
  of Technology, the University of California, and NASA. The
  Observatory was made possible by the generous financial support of
  the W. M. Keck Foundation.}

\altaffiltext{1}{Kavli Institute for Astrophysics and Space Research
  and Department of Physics, Massachusetts Institute of Technology,
  Cambridge, MA 02139, USA}

\altaffiltext{2}{Observatories of the Carnegie Institution of
  Washington, Pasadena, CA 91101, USA}

\altaffiltext{3}{Center for Galaxy Evolution, Department of Physics and Astronomy,
University of California, Irvine, CA 92697, USA}

\altaffiltext{4}{Center for Galaxy Evolution Fellow}

\begin{abstract}
We present Magellan/MIKE and Keck/HIRES high-resolution spectra of six
red giant stars in the dwarf galaxy Segue\,1. Including one additional
Segue\,1 star observed by \citeauthor{norris10_seg}, high-resolution
spectra have now been obtained for every red giant in Segue\,1.
Remarkably, three of these seven stars have metallicities below
$\mbox{[Fe/H]} = -3.5$, suggesting that Segue\,1 is the least
chemically evolved galaxy known.  We confirm previous
medium-resolution analyses demonstrating that Segue\,1 stars span a
metallicity range of more than 2\,dex, from $\mbox{[Fe/H]} = -1.4$ to
$\mbox{[Fe/H]} = -3.8$.  All of the Segue\,1 stars are
$\alpha$-enhanced, with $\mbox{[$\alpha$/Fe]}\sim 0.5$.  High
$\alpha$-element abundances are typical for metal-poor stars, but in
every previously studied galaxy [$\alpha$/Fe] declines for more
metal-rich stars, which is typically interpreted as iron enrichment
from supernova Ia.  The absence of this signature in
Segue\,1 indicates that it was enriched exclusively by massive stars.
Other light element abundance ratios in Segue\,1, including
carbon-enhancement in the three most metal-poor stars, closely
resemble those of metal-poor halo stars.  Finally, we classify the
most metal-rich star as a CH star given its large overabundances
of carbon and s-process elements.  The other six stars show remarkably
low neutron-capture element abundances of $\mbox{[Sr/H]} < -4.9$ and
$\mbox{[Ba/H]} < -4.2$, which are comparable to the lowest levels ever
detected in halo stars.  This suggests minimal neutron-capture
enrichment, perhaps limited to a single r-process or weak s-process
synthesizing event. Altogether, the chemical abundances of Segue\,1
indicate no substantial chemical evolution, supporting the idea that
it may be a surviving first galaxy that experienced only one burst of
star formation.
\end{abstract}


\section{Introduction}

The early phases of the chemical evolution of the Universe can be
reconstructed through the study of metal-poor stars.  Given their low
metallicity, these stars are assumed to have formed in the early
Universe. Since then, these long-lived stars have locked up
information on the properties of their birth gas clouds and the local
chemical and physical conditions in their atmospheres that can be
extracted through spectroscopic analysis.

Metal-poor stars in the halo of the Milky Way have thus been used for
decades to unravel the chemical enrichment history of the Galaxy
(e.g., \citealt{McWilliametal, ARAA, psss}).  In this way the Galaxy
has been found to be chemically diverse, with multiple populations and
components, and to contain a variety of substructure. These signatures
clearly show how closely the chemical evolution of a galaxy is
connected to its assembly history. But it is difficult to cleanly
uncover the various astrophysical processes that have been involved in
element nucleosynthesis and star formation in the Milky Way over
billions of years. Dwarf galaxies, however, being smaller systems with
presumably simpler formation histories, provide the means to study
chemical enrichment in a more straightforward way. At the same time,
comparison of their population of metal-poor stars with those in
  the Milky Way provides insight into the assembly of the Galactic
  halo.

In the last decade, the Sloan Digital Sky Survey (SDSS) transformed
our picture of the Milky Way's satellite galaxy population.  The wide
sky coverage and sufficiently deep photometry revealed dwarf galaxies
with total luminosities ranging from 300 to $10^5~L_{\sun}$
\citep[e.g.,][]{willman05,zucker06,belokurov07,martin08}.  These new
galaxies were found not only to be unprecedentedly faint but also
unprecedentedly dominated by dark matter \citep{sg07}.  Nonetheless,
they form a continuous sequence of stellar mass with the classical
dwarf spheroidal galaxies in terms of star formation history
\citep{brown12}, structural properties \citep{okamoto12}, and metal
content \citep{kirby08,kirby11b}.

A significant challenge to study individual stars in these dwarf galaxies
in great detail is the difficulty in achieving the spectral data
quality required for chemical abundance analyses (e.g.,
\citealt{koch_her,ufs,leo4}). Most dwarf galaxies orbit dozens of kpc
away in the outer halo and their brightest stars therefore have
typical apparent magnitudes of ${\rm V} = 17$ to 18.  However,
high-resolution spectroscopy is required to investigate the detailed
stellar abundance patterns that reflect various enrichment events, and
such spectra can only be obtained for faint stars with long
integrations on the largest telescopes available.  
Accordingly, relatively small numbers of stars in most of the
classical dwarf spheroidal (dSph) galaxies with $10^{5}\,L_{\odot}
\lesssim L \lesssim 10^{7}\,L_{\odot}$ and the ultra-faint dwarfs ($L
\lesssim 10^{5}\,L_{\odot}$) have been observed at high spectral
resolution.

\subsection{The chemical evolution of dwarf galaxies and the Milky Way}

In the Milky Way it is well known that low-metallicity halo stars
($\mbox{[Fe/H]}<-1.0$) have enhanced abundances of the
$\alpha$-elements (e.g., \citealt{edvardsson93}), while at higher
metallicities the [$\alpha$/Fe] ratios smoothly approach the solar
ratio \citep[e.g.,][]{tinsley79,mcwilliam97,venn04}.  The canonical
interpretation of this behavior is that it reflects the timescales of
nucleosynthesis from different kinds of supernovae.  Massive stars
produce large amounts of the $\alpha$-elements both during their
stellar evolution and in their explosions as core collapse supernovae.
These elements are quickly returned to the interstellar medium because
the lifetimes of such stars are short, less than 10\,Myr.  Type Ia
supernovae, which primarily produce iron, do not begin exploding until
a poorly known delay time (typically assumed to be of order
$10^{8}$\,yr; \citealt{maoz12}) has elapsed since an episode of star
formation.  The [$\alpha$/Fe] plateau at $\mbox{[Fe/H]}<-1.0$ then
corresponds to the epoch when only core-collapse supernovae
contributed significantly to the overall nucleosynthesis, and the
decline to $\mbox{[$\alpha$/Fe]} = 0$ occurs when Type\,Ia supernovae
begin occurring in significant numbers as well.

The turnover from the high [$\alpha$/Fe] plateau in the classical dSph
galaxies occurs at lower metallicity than in the Milky Way, [Fe/H]$
\sim -2.5$ (e.g., \citealt{tolstoy03, venn04}), revealing that they
have been enriched on slower timescales than what is observed in the
halo of the Galaxy (see e.g., \citealt{tolstoy_araa} for a review).
Note that, at least in the case of Sagittarius, \citet{mcwilliam13}
have questioned whether this explanation for the decline in
[$\alpha$/Fe] at high metallicity is correct, and they suggest a
top-light initial mass function (IMF) as an alternative.

More recently, \citet{kirby08,kirby11} measured metallicities for
dozens of individual stars in ultra-faint dwarfs using the
medium-resolution spectroscopy from \citet{sg07}.  Stars with
metallicities of $\mbox{[Fe/H]}<-3.0$ were uncovered in surprisingly
large relative numbers, whereas essentially no stars with
$\mbox{[Fe/H]}>-1.0$ were found.  This general characteristic of
overall metal-deficiency correlates with the low luminosities of these
galaxies. They extend the metallicity-luminosity relationship for dSph
galaxies by several orders of magnitude in luminosity
\citep{kirby08,kirby10,kirby13}.  \citet{vargas13} then used the same
spectra to measure [$\alpha$/Fe] abundance ratios.  They found that among
their sample of eight ultra-faint dwarfs, only Segue\,1 does not show
declining [$\alpha$/Fe] ratios with increasing
metallicity\footnote{Figure 4 in \citet{vargas13} indicates that
  UMa~II contains a single metal-rich ($\mbox{[Fe/H]} \sim -1.0$) and
  $\alpha$-enhanced ($\mbox{[$\alpha$/H]} \sim 0.4$) star that might
  place it in this category as well.  However, high-resolution
  spectroscopy of this star by \citet{ufs} showed that it is actually
  a foreground star rather than a member of UMa\,II.}.  Thus, this
galaxy is the only dwarf galaxy known to have minimal chemical
enrichment from Type\,Ia supernovae.

However, medium-resolution spectroscopy is limited in its ability to
detect and measure the abundances of some elements, such as the
neutron-capture elements.  High-resolution spectroscopy can provide
highly detailed abundance information for stars that are bright
enough.  High-resolution spectra with large wavelength coverage of
stars down to magnitude of $V\sim19.2$ have been obtained for about a
dozen stars in ultra-faint dwarfs by now. Three stars each in Ursa
Major\,II and Coma Berenices were observed by \citet{ufs},
\citet{koch_her} observed two stars in Hercules, and other studies
reported on one star each in Leo\,IV \citep{leo4}, Bo\"otes\,I
\citep{norris10}, Segue\,1 \citep{norris10_seg}, and the stream
passing just in front of Segue\,1 that may have originated in an
ultra-faint dwarf \citep{frebel13b}.

Detailed studies of these stars, nearly all of which are at
$\mbox{[Fe/H]}\lesssim-2.0$, revealed close-to-identical chemical
abundance patterns compared with halo stars, both in terms of the
abundance ratios as well as more global population signatures such as
a significant fraction of metal-poor stars being strongly enhanced in
carbon relative to iron, and showing low neutron-capture element
abundances. It has thus been suggested that the chemical similarity of
halo and the ultra-faint dwarf galaxy stars could be due to the stars
having formed from early gas that was enriched in the same fashion,
i.e., exclusively by massive stars \citep{ufs, leo4, norris10,
  norris10_seg}. Moreover, if the surviving ultra-faint dwarf galaxies
had earlier analogs that were accreted by the Milky Way in its early
assembly phases, the chemical similarity of halo and dwarf galaxy
stars could also be interpreted as indicating that the early
enrichment history of the destroyed dwarfs closely resembled that of
the surviving dwarfs observed today, despite the different
environments they formed in.  In this picture, very low luminosity
primordial dwarfs may have provided the now-observed metal-poor
``halo'' stars to the halo.

In this context, it is interesting to note that the analysis of deep
\emph{HST} color-magnitude diagrams of three ultra-faint dwarf
galaxies, Hercules, Leo\,IV, and Ursa Major\,I, by \citet{brown12}
shows that they are at least as old as the oldest globular clusters
and likely nearly as old as the Universe itself (their age is
consistent with that of the globular cluster M92, which on the same
scale is measured at 13.7\,Gyr).  Preliminary results for three
additional ultra-faint dwarfs suggest that the stellar populations of
all six galaxies are indistinguishable \citep{brown13}.  This
demonstrates that the metal-poor stars in these galaxies are as old as
their chemical composition (galaxy average metallicities range from
$\mbox{[Fe/H]}\sim-2.0$ to $-2.6$; \citealt{kirby08}) suggests.
Because of the apparently small age spreads in these systems, their
more metal-rich stars also have to be similarly old, suggesting rapid
enrichment. This could be due to the low-mass nature of these systems
which would lead to a fast, significant build up of metals after only
a few supernova explosions. Considering a plausible chemical composition
of a first galaxy that may have survived to the present day,
\citet{frebel12} thus argued that Segue\,1 (together with Ursa
Major\,II, Coma Berenices, Bo\"otes\,I, and Leo\,IV) are candidate
systems for such surviving first galaxies. \citet{bovill11} agree that
most of the ultra-faint dwarfs are consistent with expectations for
reionization fossils, although they do not place Segue~1 in this
category as a result of earlier estimates of its metallicity lying
above the luminosity-metallicity relation established by brighter
dwarfs.  The metallicities derived in this paper and improved
measurements of the L--Z relation by \citet{kirby13} demonstrate that
Segue~1 is in fact consistent with the extrapolated
metallicity-luminosity relationship of more luminous systems.  Only
additional chemical abundance data for more stars in as many of the
ultra-faint dwarfs as possible will allow detailed tests of the
hypothesis that these objects are fossils of the first galaxies by
establishing a detailed account of the chemical composition of each
galaxy.

Thus, in this study we present chemical abundance measurements
  for six stars in Segue~1 that are just bright enough to be
  observable with high-resolution spectroscopy. Segue~1 is the
  faintest galaxy yet detected, and it has an average metallicity of
  $\mbox{[Fe/H]} \sim -2.5$ to $-2.7$ \citep{norris10_booseg,
    simon11}.  It was one of five new ultra-faint galaxies discovered
  by \citet{belokurov07} using a matched filter search of SDSS DR5 and
  SEGUE photometry.  Although Segue~1 contains very few stars, its
  position $50\degr$ out of the Galactic plane and away from the
  Galactic center aids in separating member stars from the foreground
  Milky Way population.  It is also close enough (23\,kpc) to permit
  spectroscopy of stars down to $\sim1$\,mag below the main sequence
  turnoff \citep[e.g.,][]{geha09,simon11}.  Segue 1 was initially
  presumed to be a globular cluster because of its small half-light
  radius (30\,pc), but \citet{geha09} presented a strong case based on
  internal stellar kinematics that Segue~1 is highly dark
  matter-dominated and therefore a galaxy.  \citet{geha09} also
  demonstrated that Segue~1 lies on or near standard dwarf galaxy
  scaling relations.  \citet{niederste09} found photometric evidence
  for tidal debris near Segue~1 and proposed that the velocity
  dispersion of Segue~1 was inflated by contamination from these
  disrupted structures, but \citet{simon11} showed that contamination
  is unlikely and that the measured velocity dispersion is robust.
  The extensive spectroscopy by both \citet{simon11} and
  \citet{norris10_booseg} also established that Segue~1 has
  metallicity spreads of 0.7 to 0.8\,dex in [Fe/H] and 1.2\,dex in
  [C/H]. Along with being extremely underluminous and the most dark
  matter-dominated and lowest-metallicity object currently known,
  Segue~1 is not only a galaxy, but perhaps the most extreme galaxy
  known.

With our new observations, our aim is to quantify the chemical
evolution of this galaxy by constraining its enrichment
processes. This way, we can learn about the limited star formation
that occurred in this early system. In \S\,\ref{sec:obs} we describe
the observations and in \S\,\ref{sec:analysis} our analysis
techniques. We interpret our chemical abundance results
(\S\,\ref{signature}) within the context of early galaxy formation and
chemical evolution in \S\,\ref{history}. We conclude in
\S\,\ref{sec:conc}.

\section{Observations}\label{sec:obs}

\subsection{Target selection}\label{selection}

The first spectroscopy of Segue\,1 was obtained by \citet{geha09}, who
used medium resolution Keck/DEIMOS spectra to identify 24 member stars
of the galaxy, including a single red giant for which they estimated
$\mbox{[Fe/H]} = -3.3$.  \citet{simon11} followed up this study by
observing an essentially complete spectroscopic sample (again with
Keck/DEIMOS) out to 2.3 half-light radii around Segue\,1 and down to a
magnitude limit of $r = 21.7$ ($\sim1$\,mag below the main sequence
turnoff).  The \citet{simon11} survey identified five additional
members on the Segue\,1 red giant branch (RGB), and a total of 71
member stars.  Independently, \citet{norris10_booseg} used medium
resolution blue spectra from AAOmega on the Australian Astronomical
Telescope (AAT) over a much wider field to search for bright Segue\,1
members.  They found four of the same RGB stars as \citet{simon11}, as
well as another extremely metal-poor (EMP) giant, Segue\,1-7, nearly
four half-light radii from the center of the galaxy.  One other
candidate RGB member from the \citet{norris10_booseg} sample,
Segue\,1-42, was found by \citet{simon11} to have a radial velocity
inconsistent with membership in Segue\,1, while the last (and
faintest) RGB candidate identified by \citeauthor{norris10_booseg},
Segue\,1-270, is unlikely to be a genuine member given its location
$\sim10$ half-light radii from Segue\,1 and its velocity offset of
$\sim20$~km~s$^{-1}$ from the systemic velocity.

As a result of these extensive observations and the remarkably puny
stellar population of the galaxy, the seven known RGB stars are likely
to represent a complete inventory of all stars in Segue\,1 currently
in the red giant phase of evolution (it also has two horizontal branch
stars; \citealt{simon11}).  For a Plummer radial profile, 10\%\ of
member stars should be located at projected radii beyond three
half-light radii, within which the \citet{simon11} sample is more than
90\%\ complete.  In fact, two of the seven giants (29\%) are beyond
this distance, suggesting that the present sample
includes most (if not all) of the RGB stars in
Segue\,1. \citet{norris10_seg} obtained a high-resolution VLT spectrum
of the brightest giant star, Segue\,1-7, and analyzed its unusual
chemical inventory, confirming its extremely low metallicity and
carbon enhancement.  As described below, we have now obtained high
resolution spectra of the remaining six Segue\,1 giants, including two
stars with $\mbox{[Fe/H]} < -3.5$ and two others at $\mbox{[Fe/H]} >
-1.8$ that are the most metal-rich stars in an ultra-faint dwarf
galaxy to be studied in detail.

\subsection{High-resolution spectroscopy}
We observed five of our six target stars with the MIKE spectrograph
\citep{mike} on the Magellan-Clay telescope in March and May 2010, and
March 2011.  Observing conditions during these runs were mostly clear,
with an average seeing of 0.8\arcsec\ to 1.0\arcsec.  Additional
details of the MIKE observations are given in
Table~\ref{Tab:obs}. MIKE spectra have nearly full optical wavelength
coverage from $\sim3500$-9000\,{\AA}.  A $1.0\arcsec\ \times 5\arcsec$
slit yields a spectral resolution of $\sim22,000$ in the red and
$\sim28,000$ in the blue wavelength regime. We used $2\times2$ on-chip
binning and the 1.0\arcsec\ slit for all stars except
SDSS\,J100639+160008. The seeing conditions were better when we began
observing this star, so we opted to employ a $0.7\arcsec\ \times
5\arcsec$ slit. It yields a resolution of $\sim28,000$ and
$\sim35,000$, respectively.

\begin{deluxetable*}{lrrcccrcccc} 
%\rotate
\tablewidth{0pt} 
\tabletypesize{\scriptsize}
\tablecaption{\label{Tab:obs} Observing Details }
\tablehead{
\colhead{Star} & \colhead{$\alpha$}&\colhead{$\delta$}&\colhead{UT dates}&\colhead{slit}&\colhead{$t_{\rm {exp}}$} 
&\colhead{$g$} &\colhead{$E(B-V)$} &\colhead{$S/N$} &\colhead{$S/N$}   \\
\colhead{}&\colhead{(J2000)}&\colhead{(J2000)}&\colhead{ }&\colhead{}&\colhead{hr}&\colhead{mag}&\colhead{mag}&\colhead{5300\,{\AA}}&\colhead{6000\,{\AA}} }
\startdata
%    star         RA           dec             UT                               slit   texp   mag    ebv   sn5300/6000    comment
 SDSS\,J100714+160154& 10~07~14.6  &+16~01~54.5 & 2011-03-10                     &$1.0\arcsec$ & 5.5& 18.86& 0.027& 35& 45\\ % segue-98 from Norris
 SDSS\,J100710+160623& 10~07~10.1  &+16~06~23.9 & 2011-03-13/22                  &$1.0\arcsec$ & 7.3& 19.20& 0.026& 28& 45\\%seg-318, but no feh 
 SDSS\,J100702+155055& 10~07~02.5  &+15~50~55.3 & 2011-03-11/12                  &$1.0\arcsec$ & 9.2& 18.50& 0.033& 35& 46\\%seg-71 -1.9 from Norris
 SDSS\,J100742+160106& 10~07~42.7  &+16~01~06.9 & 2010-04-01                     &$1.15\arcsec$& 3.6& 18.60& 0.027& 20& 30\\%seg-31 -2.2 from Norris
 SDSS\,J100652+160235& 10~06~52.3  &+16~02~35.8 & 2010-03-07/08/18/19/23/24      &$1.0\arcsec$ & 15 & 18.89& 0.029& 42& 50\\
 SDSS\,J100639+160008& 10~06~39.3  &+16~00~08.9 & 2010-03-18/19/22, 2010-05-08/09&$0.7\arcsec$ & 8  & 19.48& 0.031& 26& 33
\enddata
\tablecomments{The $S/N$ is measured per $\sim33$\,m{\AA} pixel (MIKE spectra) and $\sim20$\,m{\AA} pixel (HIRES spectrum).}
\end{deluxetable*} 

Integration times ranged from $\sim6$ to 15\,h.  The observations were
typically broken up in 55\,min exposures to avoid significant
degradation of the spectra by cosmic rays. These individual spectra
generally had low counts given the faintness of the objects, with
little or no flux detected below 4000\,{\AA}.

\begin{figure*}[!t]
 \begin{center}
  \includegraphics[clip=true,width=8cm, bbllx=42, bblly=123,
    bburx=558, bbury=718]{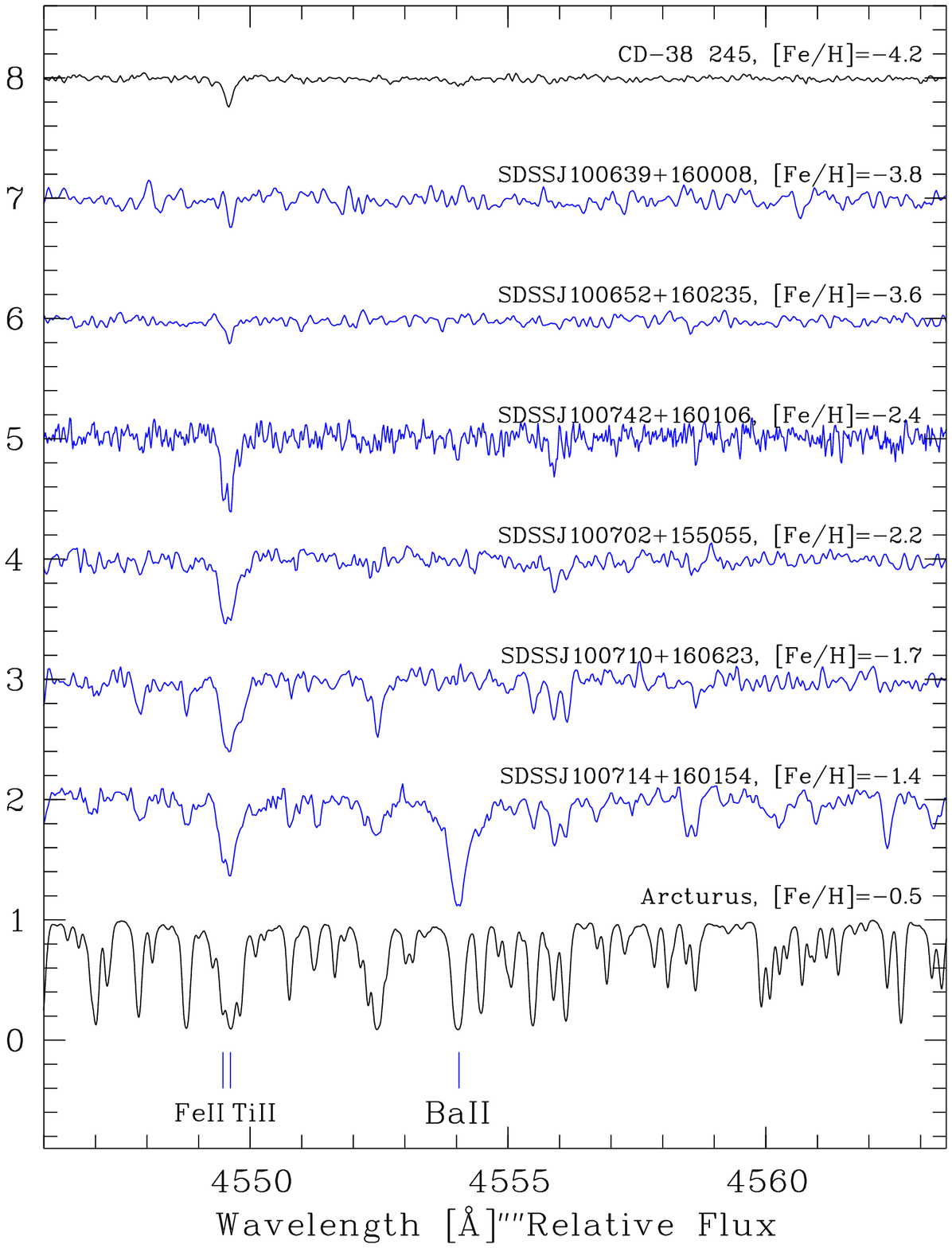}
  \includegraphics[clip=true,width=8cm, bbllx=44, bblly=124,
    bburx=558, bbury=718]{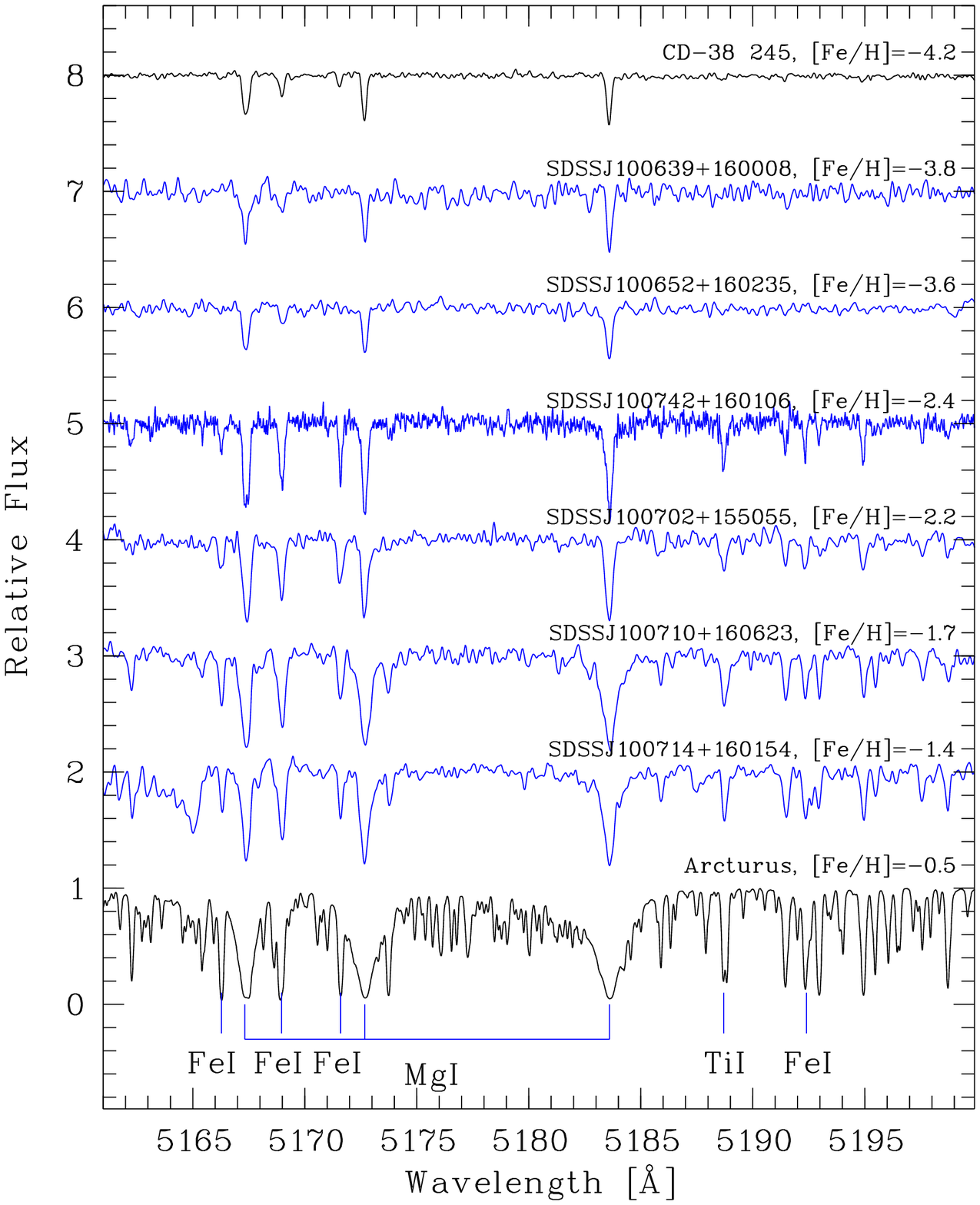}
  \figcaption{\label{specs}Magellan/MIKE and Keck/HIRES spectra of our
    Segue\,1 sample stars, shown near the Ba\,II line at 4554\,{\AA}
    (left panel) and near the Mg\,b lines around 5180\,{\AA} (right
    panel).  Some absorption lines are indicated.  The stars are
    bracketed in terms of their metallicity by the Arcturus (bottom) and
    CD~$-$38~245 (top) spectra, to illustrate the large metallicity spread in
    Segue\,1. The Ba\,II line is only detected in two stars,
    indicating that Segue\,1 is deficient in neutron-capture elements.}
 \end{center}
\end{figure*}

We observed the final star in the sample, SDSS\,J100742+160106, with the HIRES spectrograph \citep{vogt94}
on the Keck\,I telescope on 2010 April 1.  The observations were
obtained with a 1.15$\arcsec \times 7\arcsec$ slit (providing a
spectral resolution of 37,500), the kv389 blocking filter, and a total
integration time of 3.6\,h.  The autoguider had difficulty guiding
accurately during much of this time, initially because of unusually
good seeing conditions and later as the star transited within a few
degrees of the zenith, so many of the exposures were either unguided
or employed manual guiding by the telescope operator.  These guiding
issues decreased the S/N of some exposures, but should not have any
effect on the measurement of equivalent widths we are interested in
for the purposes of this paper.  We reduced the HIRES spectra using
the IDL pipeline developed by J.~X. Prochaska and
collaborators\footnote{http://www.ucolick.org/~xavier/HIRedux/index.html}.

Reductions of the individual MIKE spectra were carried out using the
MIKE Carnegie Python pipeline initially described by \citet{kelson03}
\footnote{Available at http://obs.carnegiescience.edu/Code/python}.
The orders of the combined spectrum were normalized and merged to
produce final one-dimensional blue and red spectra for further
analysis. The $S/N$ of the spectra is modest and ranges from 20 to 40
at $\sim5300$\,{\AA} and 30 to 50 at $\sim6000$\,{\AA}. Radial
velocity measurements yield values between 200 and 208\,km\,s$^{-1}$,
which are consistent with previous measurements \citep{simon11}. 

In Figure~\ref{specs} we show representative portions of the spectra
of the program stars around the Ba line at 4554\,{\AA} and the Mg\,b
lines at 5170\,{\AA}. The large range of metallicities found in
Segue\,1 is easily visible. For comparison we also add CD~$-$38~245
with $\mbox{[Fe/H]}\sim-4.2$ as well as Arcturus
($\mbox{[Fe/H]}=-0.5$), which bracket the metallicities of our
Segue\,1 stars.

\section{Chemical abundance analysis}\label{sec:analysis}

\subsection{Line measurements}
We measured the equivalent widths of metal absorption lines throughout
the spectra by fitting Gaussian profiles to them. Continuum placement
was challenging at times given the modest $S/N$ of the
data. Generally, lines between 4000\,{\AA} and 7000\,{\AA} were
measured, based on a linelist described in \citet{cash1}. In
Table~\ref{Tab:Eqw}, we list the lines used and their measured
equivalent widths for all elements, and 3$\sigma$ upper limits for
selected elements.  For blended lines, lines with hyperfine-structure
(HFS), and molecular features such as CH, we used the spectral
synthesis approach in which the abundance of a given species is
obtained by matching the observed spectrum to a synthetic spectrum of
known abundance.

\input{equivalent_widths_segue_stars}

In many instances the element abundances were so low that no
absorption lines could be detected. The modest $S/N$ of the data also
made detections of weak lines difficult.  This was the case for all
neutron-capture elements in all our stars except the most metal-rich
object and the Ba measurements in SDSS\,J100742+160106 (see
  Section~\ref{ncaps}). We determined upper limits through spectrum
synthesis by matching synthetic spectra to the noise level in the
region of the non-detected absorption line. These generally agree well
with calculations for 3$\sigma$ upper limits.

\subsection{Stellar parameters}\label{stell_par_description}

We use 1D plane-parallel model atmospheres with $\alpha$-enhancement
from \citet{castelli_kurucz} and the latest version of the MOOG
analysis code \citep{moog, sobeck11}. The abundances are computed
under the assumption of local thermodynamic equilibrium (LTE). We
calculated final abundance ratios [X/Fe] using the solar abundances of
\citet{asplund09}. The elemental abundances for our sample are given
in Table~\ref{abund}.

To derive stellar parameters spectroscopically we follow the
  procedure described in \citet{frebel13}. For completeness we repeat
  essential details below. We derive spectroscopic effective
  temperatures by demanding that there be no trend of \ion{Fe}{1} line
  abundances with excitation potential. We use initial temperature
  guesses based on the shape of the Balmer lines. Using the ionization
  balance, i.e., demanding that \ion{Fe}{1} lines yield the same
  abundance as \ion{Fe}{2} lines, we derive the surface gravity, $\log
  g$, for all sample stars. The microturbulence, \mbox{v$_{\rm
      micr}$}, is obtained iteratively in this process by demanding no
  trend of abundances with reduced equivalent widths.

However, this spectroscopic technique is known to deliver cooler
temperatures (and lower surface gravities) than those derived from
photometry, especially for cool giants (see e.g., \citealt{hollek11}
for a discussion). To alleviate this effect, we thus applied
temperature adjustments of $T_{\rm{eff, corrected}} = T_{\rm{eff,
    initial}} - 0.1 \times T_{\rm{eff, initial}} + 670$ as presented
in \citet{frebel13} after obtaining the spectroscopic parameters. This
relation was derived from the analysis of seven well-studied stars in
the literature for which photometric temperatures were available. We
then repeated determining the stellar parameters spectroscopically but
with the adjusted temperature fixed while finding the new,
corresponding $\log g$ and \mbox{v$_{\rm micr}$}.  We estimate our
temperature uncertainties to be $\sim100 - 150$\,K (see also
\citealt{ufs} as well as Section~\ref{unc} for more details on
uncertainties).  Uncertainties in $\log g$ and \mbox{v$_{\rm micr}$}
are estimated to be 0.3\,dex and 0.3\,km\,s$^{-1}$. Finally, we
visually inspect the shape of the Balmer lines to confirm the
temperatures. Compared to other stars with known temperatures, our
derived values are in qualitative agreement with the shapes of the
Balmer lines for those temperatures. Having available stellar
temperatures that are not as discrepant from photometric temperatures
makes subsequent abundance comparisons with other stars more
straightforward. Another advantage of the spectroscopic approach over
photometric temperatures is that it is reddening-independent (although
Segue\,1 has only moderate reddening of $E(B-V)\sim0.03$\,mag).

\begin{figure}[!th]
 \begin{center}
   \includegraphics[clip=true,width=8cm,bbllx=67, bblly=470,
     bburx=500, bbury=786]{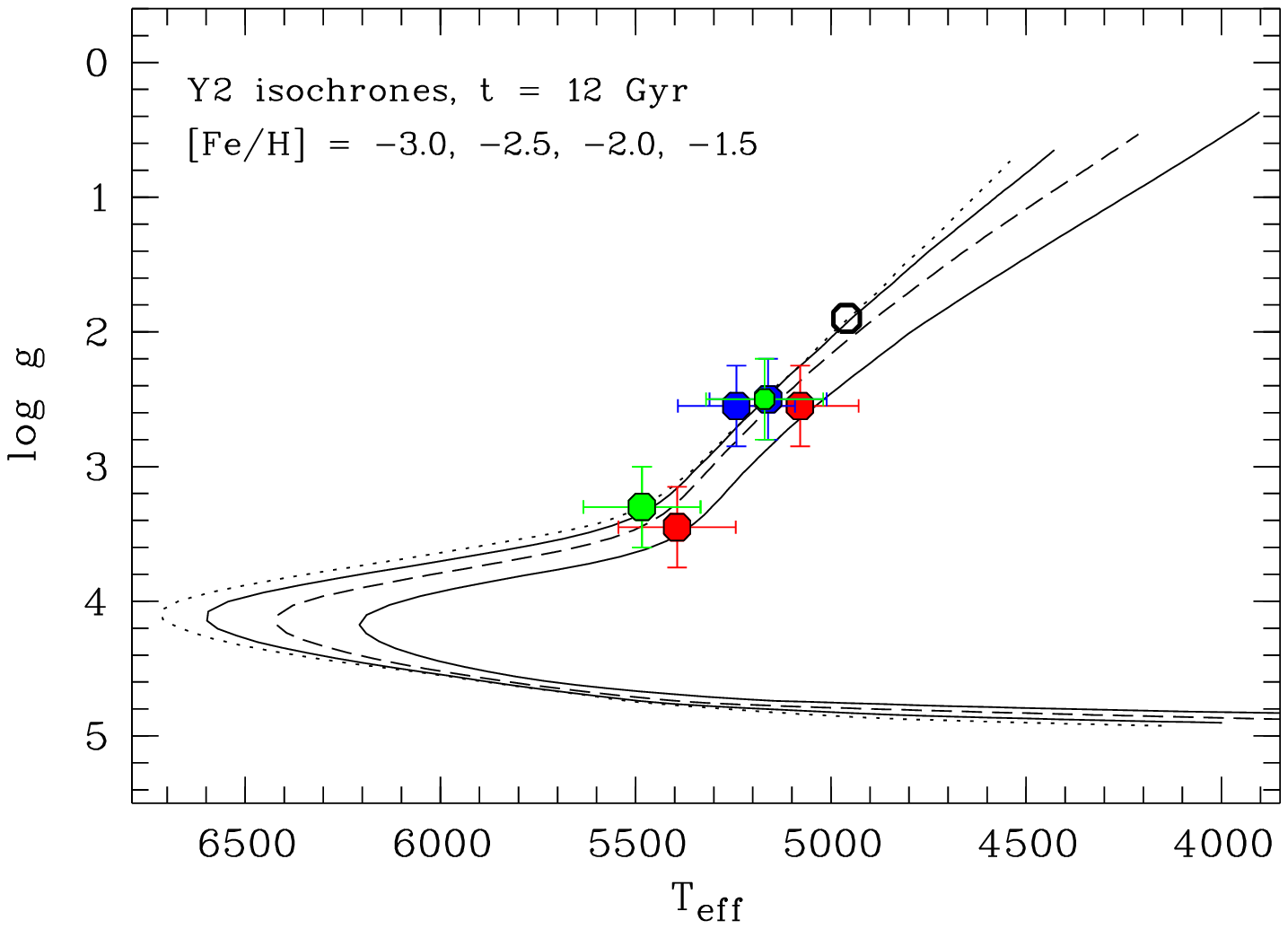}
   \figcaption{ \label{iso} Final stellar parameters in comparison
     with 12\,Gyr isochrones with $\mbox{[$\alpha$/Fe]}=0.4$ and
     metallicity ranging from $\mbox{[Fe/H]} = -1.5$ to $-3.0$
     \citep{green, Y2_iso}. Green circles indicate stars with
     $\mbox{[Fe/H]} < -3.0$, blue circles $-3.0 < \mbox{[Fe/H]} <
     -2.0$, red circles $\mbox{[Fe/H]} > -2.0$. The black circle shows
     Segue\,1-7 from \citet{norris10_seg}.}
 \end{center}
\end{figure}

Figure~\ref{iso} shows our final stellar parameters in comparison with
$\alpha$-enhanced ($\mbox{[$\alpha$/Fe]}=0.3$) 12\,Gyr isochrones
\citep{green, Y2_iso} covering a range of metallicities. Our values
agree very well with those of the isochrone. Table~\ref{Tab:stellpar}
lists the individual stellar parameters of all stars. For completeness
we added Segue\,1-7, the other star in Segue\,1 that was observed at
high spectral resolution by \citet{norris10_seg}.

Given the modest $S/N$ of the spectra, we note that relatively few
iron lines are measured, especially for Fe\,II. In the case of the two
most metal-poor stars, no Fe\,II lines for the gravity determination
were confidently detected.  We thus chose gravity values from the
isochrone appropriate for their effective temperatures. We note,
however, that upper limits on the Fe\,II lines (taken at face value)
would yield gravities within 0.5\,dex of the adopted values.  This
suggests that our final surface gravities, which place them on the red
giant branch rather than the horizontal branch, are reasonable choices
(see also below).

We also used published equivalent width measurements from
\citet{norris10_seg} to obtain stellar parameters and abundances for
Segue\,1-7 in the same way as for our other Segue\,1 stars. The
agreement with the values they determined is very good. We find
T$_{\rm eff}=4990$\,K, $\log g=2.05$, $v_{micr}=1.35$\,km\,s$^{-1}$
and $\mbox{[Fe/H]}=-3.55$, compared to 4960\,K, 1.9, 1.3\,km\,s$^{-1}$
and $-3.57$ \citep{norris10_seg}.

\begin{figure*}[!tb]
 \begin{center}
   \includegraphics[clip=true,width=15cm]{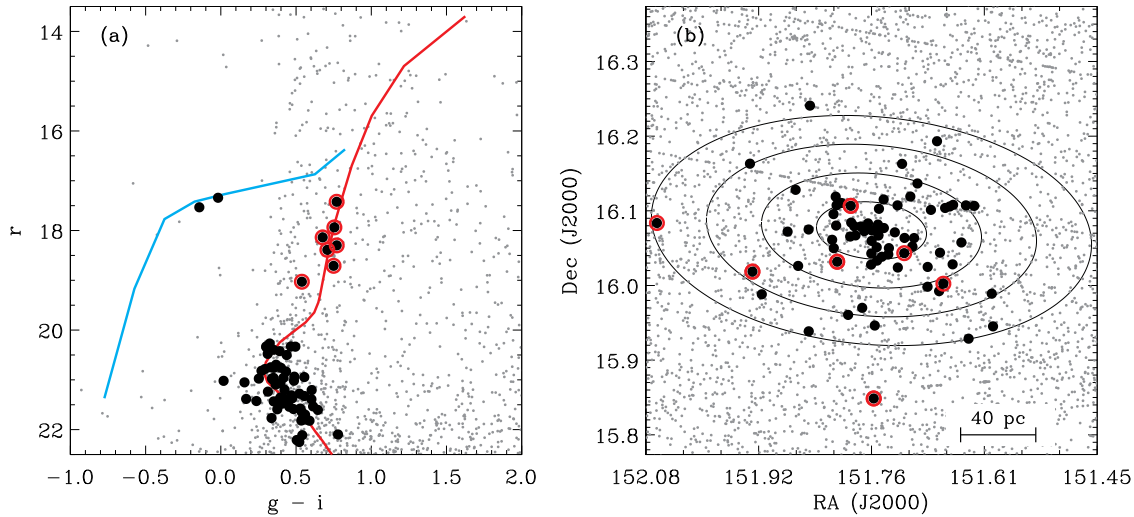}
   \figcaption{ \label{cmd} Color-magnitude diagram of Segue\,1 (left
     panel) together with the spatial distribution of the stars (right
     panel).  In the left panel, the black points are the 71 Segue\,1
     members identified by Simon et al. (2011) plus Segue\,1-7 from
     \citet{norris10_seg}. The stars outlined in red are the 7 giants
     discussed in this paper, and the gray dots are all stars from the
     SDSS catalog in the area shown in the right panel.  The RGB/main
     sequence fiducial track is the slightly modified M92 sequence
     from Simon et al. (2011; originally taken from \citealt{clem08}),
     and the HB track is from M13. The symbols are the same in the
     right panel, and ellipses marking 1, 2, 3, and 4 half-light radii
     are overplotted.}
 \end{center}
\end{figure*}

We show the available SDSS photometry in Figure~\ref{cmd} in the form
of a color-magnitude diagram overlaid with an M92 fiducial sequence
and a HB track from M13. g magnitudes and reddening are also listed in
Table~\ref{Tab:obs}. Overall, the location of the stars on the giant
branch in the photometry is confirmed with the stellar spectroscopic
parameters. We note that the $g-i$ color of SDSS\,J100639+160008 from
the Sloan Digital Sky Survey is slightly offset from the best-fitting
Segue 1 isochrone by $\sim0.1$\,mag to the blue; since this offset
disappears in other colors (e.g., $g-r$ and $r-z$), it could indicate
an error in the SDSS $i$ magnitude.  Since this star has the lowest
metallicity in the sample we were not able to determine its surface
gravity spectroscopically because no Fe\,II lines were detected.
However, we explicitly checked whether we could have obtained a
slightly lower than adopted gravity, but the upper limit of Fe\,II
lines suggest a gravity close to the isochrone or perhaps slightly
higher.  The spectrum is not consistent with a gravity high enough to
place the star on the main sequence.  Even with a higher gravity the
metallicity does not increase significantly, ruling out the
possibility that SDSS\,J100639+160008 is a more metal-rich foreground
star. We furthermore show the spatial distribution of the full sample
of Segue\,1 members from \citet{simon11} in Figure~\ref{cmd}, along
with the locations of the stars studied in this paper, and also
Segue\,1-7 from \citet{norris10_seg} which is located near 4
half-light radii, and hence was not covered in the \citet{simon11}
study.

To investigate the possibility of contamination of the Segue\,1 giant
sample by foreground stars, we consulted the Besan{\c c}on model of
the Milky Way \citep{robin03}.  The model predicts that the surface
density of Milky Way stars that meet the color and velocity cuts used
by \citet{simon11} to identify Segue\,1 members and have magnitudes
consistent with the Segue\,1 giants ($19.5 > r > 17$) is
0.8\,stars\,deg$^{-2}$, or 0.07 stars in the area over which the
\citet{simon11} survey is essentially complete.  This estimate is in
agreement with the empirical sample of non-Segue\,1 members from
\citet{simon11}.  Moreover, $\sim90$\% of these potential contaminants
are actually on the main sequence, and so would not be confused with
Segue\,1 giants given the surface gravity measurements from our
high-resolution spectra.  Thus, we expect only $\sim0.01$ Milky Way
halo giants with the same color, magnitude, and velocity as Segue\,1
to be spatially coincident with the galaxy; since there are only 7
Segue\,1 giants, it is very unlikely that our sample contains a Milky
Way star masquerading as a Segue\,1 member.

\begin{deluxetable}{lcccc} 
\tablecolumns{7} 
\tablewidth{0pt} 
\tabletypesize{\small}
\tablecaption{\label{Tab:stellpar} Stellar Parameters}
\tablehead{
\colhead{Star}  &
\colhead{$T_{\rm{eff}}$ } & 
\colhead{$\log (g)$ }    & 
\colhead{$\mbox{[Fe/H]}$ }  & 
\colhead{$v_{\rm{micr}}$ }  \\
\colhead{}&
\colhead{[K]}&
\colhead{[dex]}&\colhead{[dex]}&\colhead{[km\,s$^{-1}$]}}
\startdata
SDSS\,J100714+160154  & 5394 & 3.45 &$-$1.42 & 1.70 \\ 
SDSS\,J100710+160623  & 5079 & 2.55 &$-$1.67 & 1.75 \\ 
SDSS\,J100702+155055  & 5161 & 2.50 &$-$2.32 & 1.80 \\ 
SDSS\,J100742+160106 & 5242 & 2.55 &$-$2.40 & 1.70 \\ 
SDSS\,J100652+160235 & 5484 & 3.30 &$-$3.60 & 1.35 \\ 
SDSS\,J100639+160008 & 5170 & 2.50 &$-$3.78 & 1.55 \\ 
Segue\,1-7 & 4960 & 1.90 &$-$3.57 & 1.30 
\enddata
\tablecomments{Segue\,1-7 values are taken from \citet{norris10_seg}.}
\end{deluxetable}

\section{Chemical abundance signature of Segue\,1}\label{signature}

Our abundance analysis was carried out in a standard way, with the
goal of producing abundance patterns for our sample stars to
characterize the chemical history of Segue\,1. The final
abundances are presented in Table~\ref{abund}. We also note that we
determined abundances for Segue\,1-7, for which equivalent widths were
published by \citet{norris10_seg}. The abundances agree within
0.05\,dex. For more general details on the element measurements as
well as their nucleosynthetic origins, we refer the reader to the
discussion in \citet{ufs}, who analyzed spectra of similar metal-poor
stars in Ursa Major\,II and Coma Berenices. However, some details are
repeated here for completeness.

\subsection{Carbon} 

\begin{figure}[!tb]
\begin{center}
\includegraphics[clip=true, width=9.5cm, bbllx=54, bblly=110,
  bburx=440, bbury=768]{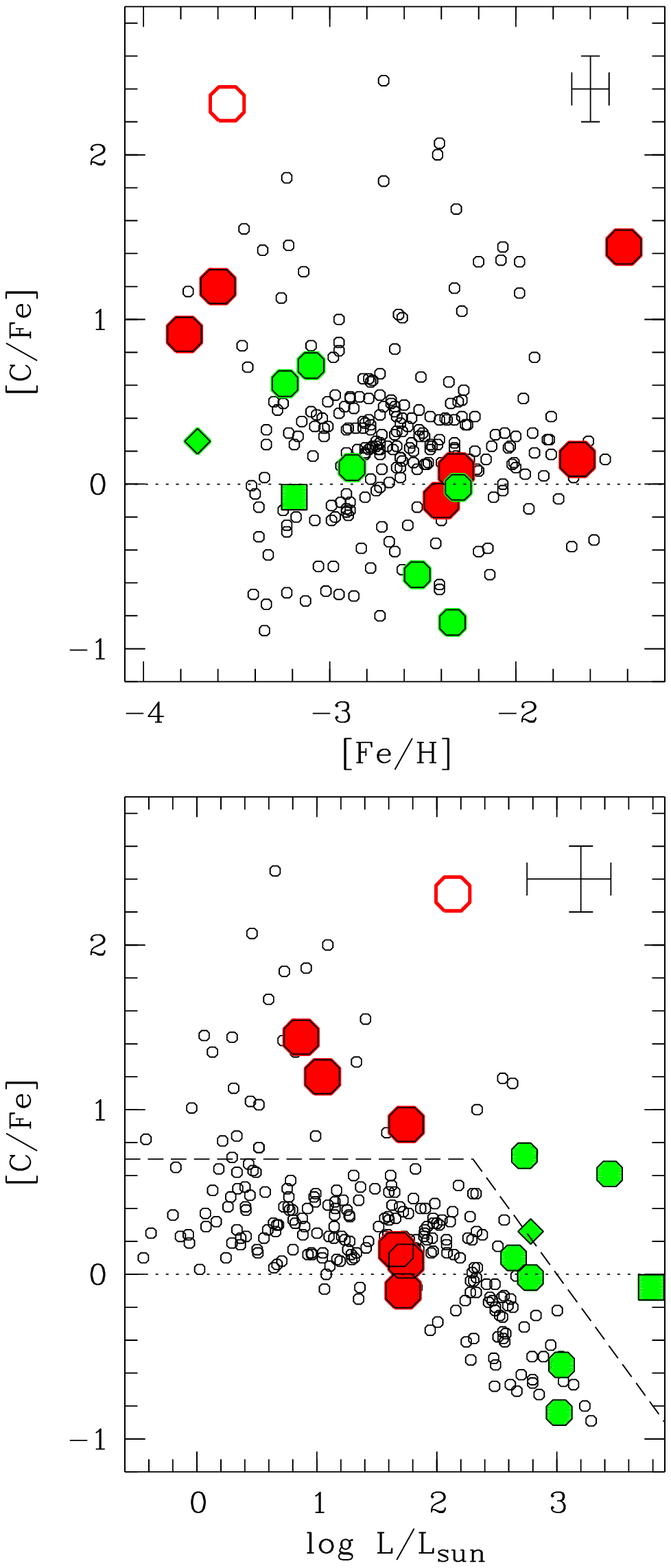} \figcaption{ \label{cfe} [C/Fe]
  abundance ratios (filled red circles) as a function of [Fe/H] (top
  panel) and stellar luminosity (bottom panel), in comparison with
  metal-poor halo stars from \citet{heresII}. The definition of
  C-enhancement from \citet{aoki_cemp_2007} is shown with a dashed
  line. Representative error bars are also shown.  The three most
  metal-poor stars (including Segue\,1-7 of \citealt{norris10_seg},
  open red circle) are carbon-enhanced. The more metal-rich stars have
  carbon abundances near the solar ratio (dotted line), with the
  exception of the CH star SDSS\,J100714+160154. Green smaller circles
  are stars in UMa\,II, ComBer \citep{ufs} and Leo\,IV \citep{leo4}.}
\end{center}
\end{figure}

Carbon is an important element for tracing early star formation as
well as enrichment and nucleosynthesis processes. We measured the
carbon abundances in all the stars from two CH features at
$\sim4313$\,{\AA} and $4323$\,{\AA}. SDSS\,J100710+160623,
SDSS\,J100702+155055 and SDSS\,J100742+160106 (with
$\mbox{[Fe/H]}=-1.7$, $-2.3$ and $-2.4$, respectively) have [C/Fe]
abundances close to the solar ratio. This is typical for many
metal-poor stars, as can be seen in Figure~\ref{cfe}. On the contrary,
SDSS\,J100714+160154 is found to possess a very large overabundance of
carbon. Given another critical abundance clue -- enhanced
neutron-capture element abundances associated with the s-process (see
below) -- this star appears to be a mildly metal-poor CH star that
received its carbon from a binary companion.
 
At $\mbox{[Fe/H]}=-3.6$ and $-3.7$, SDSS\,J100652+160235 and
SDSS\,J100639+160008 have large [C/Fe] values of 0.9 and 1.2,
respectively. Together with Segue\,1-7, having $\mbox{[C/Fe]}=2.3$ at
$\mbox{[Fe/H]}=-3.6$, they can be classified as carbon-enhanced
metal-poor (CEMP) stars, as their carbon exceeds a threshold value of
$\mbox{[C/Fe]}=0.7$ \citep{aoki_cemp_2007}. With the three most
metal-poor stars in Segue\,1 being CEMP stars, and keeping in mind
that our Segue\,1 sample is small, $\sim50$\% of Segue\,1's brightest
stars presumably formed from gas that was enriched in carbon.  (Here
we are excluding SDSS\,J100714+160154 since its carbon abundance is an
obvious result of a mass transfer event.) Moreover, this also implies
that the CEMP fraction is $\gtrsim 50\%$ for stars with
$\mbox{[Fe/H]}<-3.5$.

The fraction of CEMP stars among metal-poor halo stars is known to
increase with decreasing metallicity (e.g.,
\citealt{1999rossicarbon,frebel_bmps, lucatello2006,cohen2006}),
pointing to the importance of carbon in the early Universe. Beginning
with 15\% of stars with $\mbox{[Fe/H]}<-2.0$ and 18\% at
$\mbox{[Fe/H]}<-2.5$, the fraction becomes 25\% among stars with
$\mbox{[Fe/H]}<-3.0$, 45\% at $\mbox{[Fe/H]}\le-3.5$ and 100\% below
$\mbox{[Fe/H]}<-5.0$ (based on literature data collected by
\citealt{frebel10}). A new study by \citet{lee_ys_13} based on hundreds
of thousands of SDSS stars with medium-resolution spectra finds
similar results. The CEMP fraction for red giants in the halo is $31\%
\pm 4\%$ at $\mbox{[Fe/H]}\le-3.0$ and $33\% \pm 11\%$ at
$\mbox{[Fe/H]}\le-3.5$. The Segue\,1 measurements suggest that
the tendency of the most metal-poor stars to show carbon enhancement
is not limited to the halo, but may also be a general feature of
dwarf galaxies.

A possible explanation for this early presence of large amounts of
carbon could be the existence of rotating massive stars. They could
have provided large amounts of CNO elements during their evolution
and/or supernova explosion (e.g., \citealt{meynet06}). Moreover,
low-mass star formation may be facilitated in carbon-rich environments
because carbon (together with oxygen\footnote{Oxygen and nitrogen
  features were not detected in our spectra.}) may have provided a
cooling channel for the primordial gas to sufficiently fragment
\citep{brommnature,dtrans}. If so, the carbon abundance of the entire
system would need to be above the critical metallicity for the
observed stars to have formed. Indeed, $\mbox{[C/H]}>-3.0$ for all
stars, which is well above $D_{\rm{trans}} = \log(10^{\rm{[C/H]}}
+0.9\times10^{\rm{[O/H]}})>-3.5$ \citep{dtrans}. 

An alternative explanation for the carbon-enhancement would be mass
transfer from an unseen intermediate mass binary companion that went
through the AGB phase. However, at least those carbon-enhanced
metal-poor stars that have been shown to be in binary systems also
show significant amounts of s-process material and only occur at
metallicities of $\mbox{[Fe/H]}\gtrsim-3.0$
(e.g. \citealt{masseron10}). The three carbon-enhanced EMP stars in
Segue\,1 are different (i.e. they can be classified as CEMP-no stars)
and have $\mbox{[Fe/H]}<-3.5$. For those stars, the binary fraction is
currently unknown \citep{masseron10}. Future radial velocity
monitoring of CEMP stars will reveal whether the binary scenario could
explain carbon-enhancement at $\mbox{[Fe/H]}<-3.5$.

Following \citet{frebel12}, an early galaxy should show large
abundance spreads in terms of [X/H], as a result of inhomogeneous mixing of
supernova metal yields (e.g., \citealt{greif11}).  For [Fe/H],
Segue\,1 shows a $>2$\,dex spread. For [C/H] this is also nearly
2\,dex (again, excluding SDSS\,J100714+160154). The stars with higher
[Fe/H], however, are not CEMP stars by the current definition because
their high [Fe/H] keeps the [C/Fe] ratio down. In considering this
criterion for an entire galaxy it also becomes clear that the
definition for stellar carbon enhancement, being based on the [C/Fe]
ratio, may be an insufficient description. Regardless, assuming
inhomogeneous mixing throughout Segue\,1 resulting in [X/H] spreads,
and keeping in mind that the carbon enrichment processes were likely
completely different and/or decoupled from that of iron, the important
conclusion is that all stars show a [C/H] abundance in excess of the
critical amount, independent of their [Fe/H]. Coming back to the large
fraction of CEMP stars observed in the halo, it is plausible that the
halo CEMP stars (and some more metal-rich stars, although they would
be very difficult to identify) formed in environments similar to that
of Segue\,1.

\subsection{$\alpha$-elements}

\begin{figure}[!tb]
 \begin{center}
  \includegraphics[clip=true,width=14cm,bbllx=35, bblly=120, bburx=520,
    bbury=740]{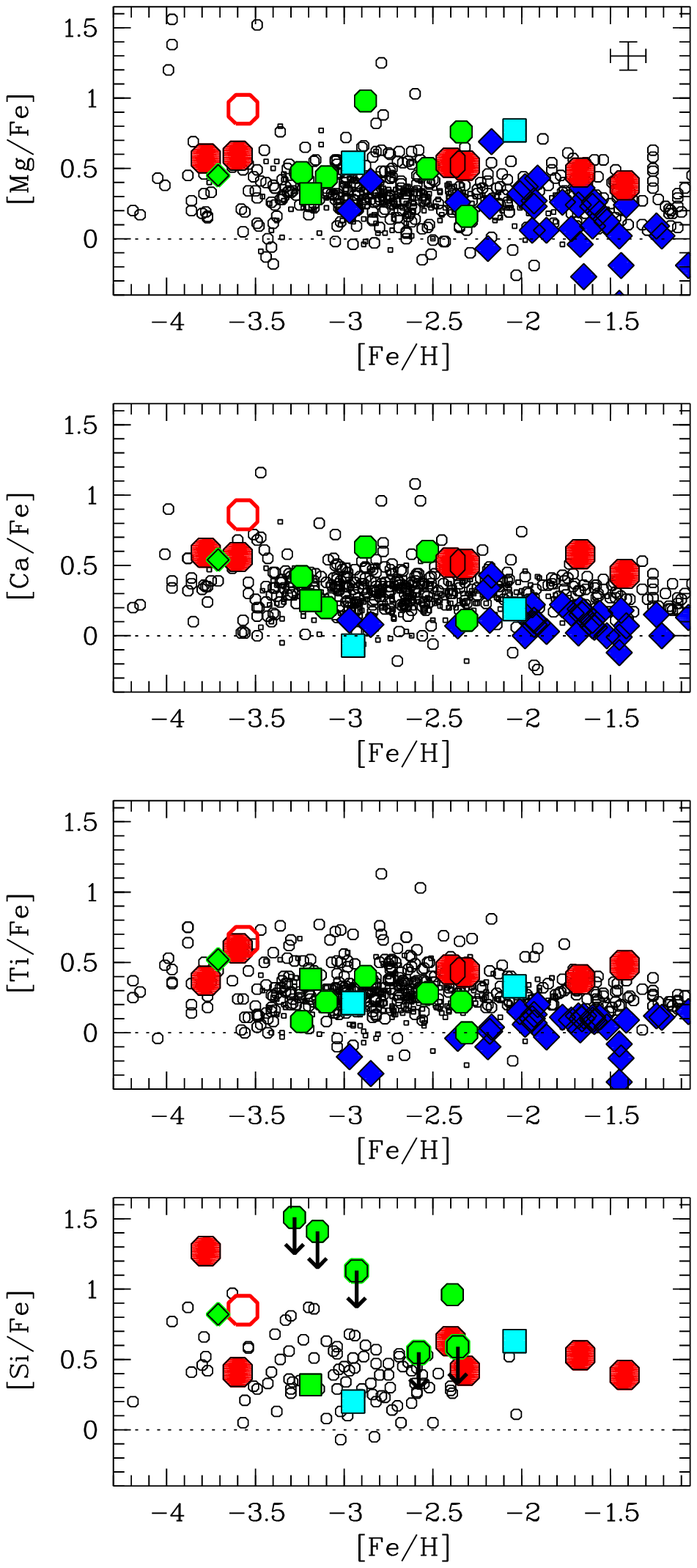} \figcaption{\label{alphas} Abundance
    ratios of $\alpha$-elements [Mg/Fe], [Ca/Fe], [Ti/Fe] and [Si/Fe]
    in our Segue\,1 stars (filled red circles) and star Segue\,1-7
    from \citet{norris10_seg} (open red circle) in comparison with
    those of other ultra-faint dwarf galaxy stars (UMa\,II, ComBer,
    Leo\,IV: green circles \citep{ufs,leo4}; Bo\"otes\,I: green
    diamond; \citep{norris10}; Hercules and Draco: cyan squares
    \citep{koch_her, fulbright_rich}), stars in the classical dwarfs
    (\textit{blue triangles}; \citealt{venn04}), and Galactic halo
    stars (black open circles; \citealt{cayrel2004, venn04, aoki05,
      heresII, yong13_II}). For Si, much less data is
    available. A representative error bars is shown in the [Ca/Fe] panel.}
\end{center}
\end{figure}

Abundances of the $\alpha$-elements magnesium, calcium and titanium
were obtained from equivalent width measurements. We used spectrum
synthesis to determine the silicon abundance. All four
$\alpha$-element abundances are enhanced at the
$\mbox{[$\alpha$/Fe]}\sim 0.5$\,dex level, even in the more metal-rich
stars including SDSS\,J100714+160154. Figure~\ref{alphas} shows the
abundance trends of all four $\alpha$-elements, in comparison with
various halo and dwarf galaxy stars. In
Figure~\ref{cayrel_abundances} we also show the comparison of all our
elemental abundances specifically with those of the extremely
metal-poor star sample from \citet{cayrel2004}. Generally, the
agreement between the samples is excellent. 

Despite the broad agreement between the abundance patterns in Segue\,1
and halo stars, careful examination of Fig.~\ref{alphas} reveals that
the Segue\,1 stars have $\sim0.1$\,dex higher abundances in each of
the $\alpha$-elements.  However, this could be due to differences in
surface gravity, $\log{gf}$ values and stellar parameter
determinations.  Similar behavior was found by \citet{hollek11}, who
traced higher Mg abundances to using systematically lower gravities
than other studies.  We tested this possibility by independently
deriving abundances for five stars in the \citeauthor{cayrel2004}
sample (HD2796, HD122563,BD $-18\deg$ 5550, CS22892-052, CS31082-001),
using their published equivalent widths and our $\log{gf}$ values and
methods of determining stellar parameters. A stellar parameter
comparison between studies can be found in
\citep{frebel13}. We find that using our technique produces no
  significant differences between our Segue\,1 $\log\epsilon
  (\mbox{X})$ abundances and the respective published
  \citeauthor{cayrel2004} abundances. The mean difference between the
  two studies for each element (O, Na, Mg, Al, Si, Ca, Sc, Ti\,I,
  Ti\,II, Cr, Mn, Fe\,I, Fe\,II, Co, Ni, Zn) ranges from $\sim$0.00 to
  $\sim$0.1\,dex with respective standard errors of the mean about as
  large or larger than the mean values. There are only two exceptions
  where the differences are more significant; there is an abundance
  difference for Na of $0.19\pm0.04$ and $0.11\pm0.015$ for
  Mg. Indeed, our Mg abundance are systematically higher by 0.1\,dex
  owing to slightly lower surface gravities. The other
  $\alpha$-elements Si, Ca and Ti, however, are not gravity sensitive
  and yield essentially identical abundances. Given the good agreement
  with the Cayrel et al. stars we conclude that the Segue\,1
$\alpha$-elements show the same behavior as what is found in typical
metal-poor halo stars with $\mbox{[Fe/H]}\lesssim-1.5$, which is
consistent with massive stars having enriched Segue\,1 at the earliest
times.

Another noteworthy issue is that the EMP star Segue\,1-7 has somewhat
higher abundance ratios than the other Segue\,1 stars. Our re-analysis
of the published equivalent widths from \citet{norris10_seg} yielded
slightly lower values, but Segue\,1-7 remains with marginally enhanced
$\alpha$-abundances relative to both the halo sample and the rest of
Segue\,1.  We also find that two of the three Segue\,1 EMP stars,
SDSS\,J100639+160008 and Segue\,1-7, show an enhancement in silicon of
$\mbox{[Si/Fe]}=1.2$ and 0.8, respectively, which is well above the
halo-typical $\alpha$-enhancement. The SDSS\,J100639+160008 Si
abundance was derived from the line at 3905\,{\AA}, which is nearly
saturated. The line is known to be blended with a CH feature, but
given the strength of the line, the contribution of CH to the observed
absorption is a minor issue. The Si abundance of Segue\,1-7 is based
on the 4102\,{\AA} line \citep{norris10_seg} and somewhat
uncertain. The similarly high Ca and Mg abundances of this star lend
support to the Si measurement.

Interestingly, though, there are other metal-poor halo stars with
$\mbox{[Fe/H]}<-2.5$ that exhibit similar behavior to these two
stars. We find 12 stars in the \citet{frebel10} database of metal-poor
stars that have $\mbox{[Si/Fe]}>0.6$ and $\mbox{[Si/Mg]}>0.4$ (four
stars from \citealt{cayrel2004}, two stars each from
\citealt{McWilliametal}, \citealt{aoki_cempno} and \citealt{lai2008},
and one star each from \citealt{preston_rhb} and
\citealt{frebel_he1300}). We use the [Si/Mg] ratio to select stars
that are not enhanced both in Si and Mg, as those exist as well. In
addition to these 12 stars, S1020549, a star with $\mbox{[Fe/H]}=-3.8$
in Sculptor \citep{scl} and one in Ursa Major\,II \citep{ufs} also
show Si abundances of $\mbox{[Si/Fe]}\sim1.0$ (although the Ursa
Major\,II star also has $\mbox{[Mg/Fe]}\sim0.7$).

These results demonstrate that high Si abundances (with and without
significant Mg overabundances) are a rather ubiquitous feature
indicating massive star enrichment driven by supernova explosions that
are perhaps sensitive to progenitor properties \citep{aoki_mg}.
Indeed, Si is made in both hydrostatic and explosive oxygen burning
processes.  A different fraction of the Si produced during stellar
evolution may survive to the SN explosion \citep{woosley_weaver_1995}
depending on the stellar mass as well as on the uncertain values for
the $^{12}$C($\alpha$,$\gamma$)$^{16}$O reaction rate.  Hence, this
could explain the relatively large spread in the observed abundances
as simply resulting from different stellar progenitor properties.

Considering all of the Segue\,1 stars, what appears most striking is
that overall, the [$\alpha$/Fe] ratios are all enhanced and nearly
identical. Moreover, there is no evolution evident with metallicity,
even above $\mbox{[Fe/H]}>-2$. The same behavior was found by
\citet{vargas13}, who used medium-resolution spectra to determine
$\alpha$-abundances for stars in many ultra-faint dwarf galaxies,
including Segue\,1. The absence of a decline in [$\alpha$/Fe] with
[Fe/H] indicates no significant contribution of Fe by Type\,Ia
supernovae to the gas clouds from which the observed stars
formed. Instead, the evidence suggests that Segue\,1 was enriched by
massive stars alone.  It is particularly notable that the highest
metallicity star in the galaxy (with $\mbox{[Fe/H]}=-1.4$) has an
average $\mbox{[$\alpha$/Fe]}=0.43$, consistent with massive star
enrichment.  For comparison, this halo-like, massive-progenitor-star
$\alpha$-enhancement is not found among metal-poor stars with
$-2.5\lesssim\mbox{[Fe/H]}\lesssim-1.5$ in the classical dwarf
galaxies. Those stars show lower $\mbox{[$\alpha$/Fe]}$ values close
to the solar value, indicating a slower enrichment timescale of their
respective hosts and a chemically ``earlier'' contribution (i.e., at
lower metallicity of the system) of iron by Type\,Ia supernovae. We
therefore conclude that chemical evolution in Segue\,1 proceeded
differently, or was truncated, in Segue\,1 compared to both the
classical dwarf spheroidals and most other ultra-faint dwarfs.

\subsection{Sodium to zinc}

\begin{figure*}[!th]
 \begin{center}
  \includegraphics[clip=true,width=16cm,bbllx=35, bblly=195, bburx=570,
   bbury=687]{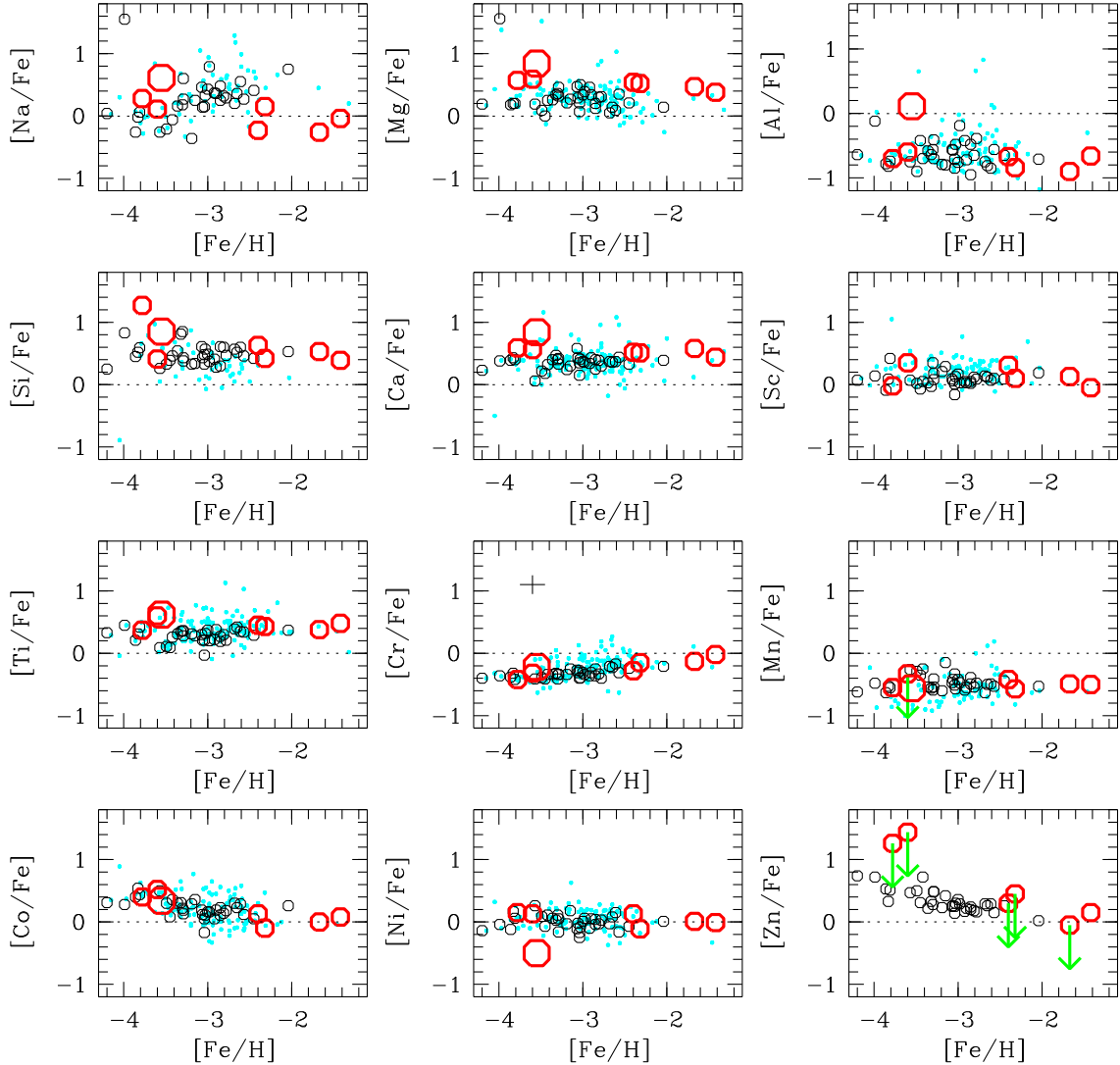} \figcaption{
     \label{cayrel_abundances} Abundance ratios ([X/Fe]) as a
     function of metallicity ([Fe/H]) for various elements detected in
     our Segue\,1 stars (small open red circles) and star Segue\,1-7
     from \citet{norris10_seg} (large red open circle) in comparison
     with those of halo stars (black circles) of \citet{cayrel2004}
     and \citet{yong13_II} (small blue points). }
 \end{center}
\end{figure*}

Various abundances of other lighter and iron-peak elements were also
determined for our Segue\,1 stars.  Overall, there is very good
agreement with the respective abundances of halo stars of similar
metallicities, as can be seen in Figure~\ref{cayrel_abundances}. Below
we briefly comment on each element.

The abundances derived from the Na\,D lines are known to be
  significantly different in a non-LTE analysis, resulting in large
  abundance corrections of 0.5\,dex for giants
  \citep{na_nlte_baum}. We note that we have not corrected our
  abundances since the literature abundances are all in LTE, and
  because our main focus is on an abundance comparison between star
  samples rather than absolute values. The sodium abundance ratios in
  our Segue\,1 stars are about that of the Sun, within $\pm
  0.3$\,dex. However, Segue\,1-7 has an even higher abundance,
  increasing the range from $\mbox{[Na/Fe]}=0.54$\,dex to 0.87\,dex.
In halo stars, aluminum is often deficient by almost 1\,dex with
respect to iron and the Sun. Our stars are no exception, with a small
scatter of $\mbox{[Al/Fe]}=0.3$\,dex. But as with Na, Segue\,1-7's
abundance is unusually high, although some halo stars have similar
abundances. 

A [Na/Fe] spread larger than that of the [Na/Al] ratio might
  be a sign of AGB nucleosynthesis and associated chemical enrichment
  as is found in globular clusters. We thus discuss Na and Al together
  to assess this possibility. Not considering Segue\,1-7, the spread
  of [Na/Fe] and [Na/Al] are both about 0.5\,dex, which is typical for
  halo stars.  When including Segue\,1-7, the [Al/Fe] spread does
  indeed become $\sim$0.2\,dex larger than the [Na/Fe]
  spread. However, globular clusters typically have high Al abundances
  of $\mbox{[Al/Fe]} > 0.4$ (e.g., \citealt{yong05, carretta12}). None
  of the seven Segue\,1 stars shares this trait. An enrichment similar
  to what is occuring within globular clusters can thus be ruled
  out. Instead, the signature of Segue\,1-7 of high C, Na, Mg, and Al
  abundances -- a pattern that has been found for some other
  metal-poor halo stars also \citep{aoki_mg} -- perhaps rather signals
  a particular kind of supernova explosion as the source of early
  metals in Segue\,1.

Scandium abundances show a tight trend in halo stars, and all Segue\,1
abundances agree extremely well with them. Chromium, manganese, cobalt
and nickel are all very similar in this regard, and again there is
excellent agreement of all Segue\,1 abundances with the respective
halo ratios. Since Mn was difficult to obtain for SDSS\,J100652+160235, we can
only present an upper limit. Segue\,1-7 has an unusually low Ni
abundance. However, it was only determined from one line
\citep{norris10_seg}. Finally, for completeness, we note that a zinc
abundance could only be determined for the most metal-rich Segue\,1
star. We show upper limits in Figure~\ref{cayrel_abundances} for all
other stars.

\subsection{Neutron-capture elements}\label{ncaps}

An emerging abundance characteristic of stars in the ultra-faint dwarf
galaxies that have been observed at high spectral resolution is that
they all display extremely low levels ($\mbox{[n-cap/Fe]}<<0.0$) of
neutron-capture elements (e.g., \citealt{koch_her,ufs,leo4,
  norris10_seg,norris10,francois12,koch13}). Segue\,1 is no exception, having the lowest
overall neutron-capture abundance level of any known system.

We attempted to measure abundances from the Sr\,II line at 4077\,{\AA}
and the Ba\,II line at 4554\,{\AA}. There is a weak Ba detection in
SDSS\,J100742+160106 and clearly visible Sr and Ba lines in
SDSS\,J100714+160154 (which is the most metal-rich star and discussed
separately below). Segue\,1-7 also has a Sr detection
\citep{norris10_seg}. For all other stars we could only derive upper
limits on these two elements. Portions of the spectra around the Ba
line can be seen in Figure~\ref{specs} (left panel). This line is
generally easily detected in metal-poor stars, even in noisy spectra,
and especially in the stars with $\mbox{[Fe/H]}>-2.5$. Hence, it is
rather unusual that it is not detected in most stars in Segue\,1. For
example, SDSS\,J100710+160623 has very low upper limits of
$\mbox{[Sr/Fe]}<-3.2$ and $\mbox{[Ba/Fe]}<-2.6$. These are the lowest
[Sr/Fe] and [Ba/Fe] values ever determined (see top panels in
Figure~\ref{ncap_plot}). However, when normalizing to hydrogen rather
than iron to assess the overall level of enrichment in Segue\,1, the
upper limits of SDSS\,J100710+160623 are at the same level as those of
the other stars (bottom panel in Figure~\ref{ncap_plot}),
$\mbox{[Sr/H]} \lesssim -5$ and $\mbox{[Ba/H]} \lesssim -4.4$.

More specific clues regarding the level at which neutron-capture
elements are present in this galaxy come from Segue\,1-7
\citep{norris10_seg} which has a measured Sr abundance of
$\mbox{[Sr/H]}=-4.9$, and from SDSS\,J100742+160106 which has a weak
Ba measurement of $\mbox{[Ba/H]}=-4.5$. However, the Ba line in
Segue\,1-7 is not detected, resulting in a limit of
$\mbox{[Ba/H]}<-4.5$. Similarly, SDSS\,J100742+160106 has an upper
limit for Sr of $\mbox{[Sr/H]}<-5.2$. The two detections, together
with the upper limits of the other stars confirm that Segue\,1, as a
system, is extremely deficient in neutron-capture elements, at or
below the 10$^{-5}$ level in [Sr/H] and 10$^{-4.5}$ level in
[Ba/H]. Nevertheless, it is apparent from the handful of Sr and Ba
detections that there is neutron-capture material present in Segue\,1,
as is the case in every other known stellar system \citep{roederer13}.

In Figure~\ref{ncap_plot}, the yellow shaded area indicates the upper
enrichment level of Sr, bounded by Segue\,1-7 at the high end.  The
same is shown for [Ba/H] with the measurement of SDSS\,J100742+160106 as the
upper bound. It is interesting to note, though, that a handful of EMP
halo stars have Sr and Ba abundances that are up to 1\,dex lower than
our upper enrichment bounds, showing that detections of these elements
at the $\mbox{[Sr/H]}\sim-6$ or $\mbox{[Ba/H]}\sim-5.5$ level are
possible, likely owing to much higher data quality.  No halo stars
above $\mbox{[Fe/H]}=-2.5$ that are so deficient in the heaviest
elements have yet been identified, demonstrating that systems like the
ultra-faint dwarfs cannot have contributed appreciably to the halo in
this metallicity range (see \S~\ref{history}).

According to \citet{simon11}, Segue\,1 has a mass within its half
light radius of $\sim6 \times 10^{5}$\,M$_{\odot}$. Its stellar mass,
however, is only $\sim10^{3}$\,M$_{\odot}$ \citep{martin08}.  Since a
typical star formation efficiency is 1\% per free fall time
\citep{krumholz07}, even if star formation in Segue\,1 only lasted for a
single free fall time, the initial gas cloud out of which the Segue\,1
stars formed was likely $\sim10^{5}$\,M$_{\odot}$.  Larger gas masses
would exceed the cosmic baryon fraction at the center of Segue\,1,
although if one considers the halo mass of the galaxy larger amounts
of baryons could be allowed.  Assuming that the gas comprising
Segue\,1 was (instantaneously and homogeneously) enriched by a first
supernova at the earliest times, an average abundance of
$\mbox{[Sr/H]}=-5.0$ (roughly that of Segue\,1-7) would imply a total
Sr mass of only $\sim10^{-7}$\,M$_{\odot}$.  However, the average Sr
abundance was likely less than that of Segue\,1-7 (as indicated by our
lowest upper limit of $\mbox{[Sr/H]}>-5.2$), which would decrease the
required total Sr mass present at Segue\,1 at early times. Hence, we
regard $\sim10^{-7}$\,M$_{\odot}$ of Sr as a reasonable estimate. In
the same way, less than $\sim10^{-7}$\,M$_{\odot}$ of Ba appears to be
present in Segue\,1, assuming $\mbox{[Ba/H]}<-4.5$.

These extremely small amounts of neutron-capture elements could have
come from just a single neutron-capture element production
event. Possibilities include the r-process occurring in a core-collapse
supernova, yielding of order $10^{-4}$\,M$_{\odot}$ of neutron-capture
elements \citep{farouqi10} or the s-process that may have operated in a
massive rotating star, perhaps a Population\,III star (e.g.,
\citealt{chiappini11, pignatari}, and see \citealt{jacobson13} for a
review on neutron-capture element sources). An AGB
enrichment event is unlikely as a source for this small amount of
neutron-capture material. Models of AGB nucleosynthesis generally
suggest much larger amounts to be produced
\citep{karakas10,lugaro12}. For example, the low-metallicity stellar
model described in \citet{placco13} loses $0.5$\,M$_{\odot}$ during
its AGB phase. This is orders of magnitude more than what is implied
by the observed low Sr and Ba abundances in the Segue\,1 stars.

More detailed yield calculations, paired with simulations of
metal-mixing in such small systems, are required to fully characterize
the neutron-capture enrichment process in systems like
Segue\,1. However, the fact that Segue\,1 has such a uniformly
low overall neutron-capture abundance suggests that the halo
stars with equally low [Sr/H] and [Ba/H] may have originated in
similar systems before being accreted by the Milky Way.

\begin{figure}[!tb]
  \begin{center}
   \includegraphics[clip=true,width=8cm,bbllx=36, bblly=340,
     bburx=535,bbury=740]{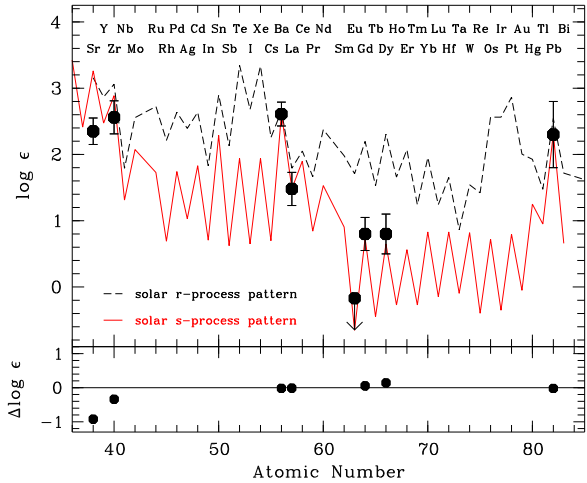}\figcaption{\label{seg100_sproc}
     Neutron-capture element abundances in SDSS\,J100714+160154 compared to the
     solar r- and s-process patterns \citep{2000burris}. SDSS\,J100714+160154
     clearly shows a signature of the s-process.}
  \end{center}
 \end{figure}

\begin{figure}[!tb]
  \begin{center}
   \includegraphics[clip=true,width=9cm,bbllx=56, bblly=223,
     bburx=535, bbury=608]{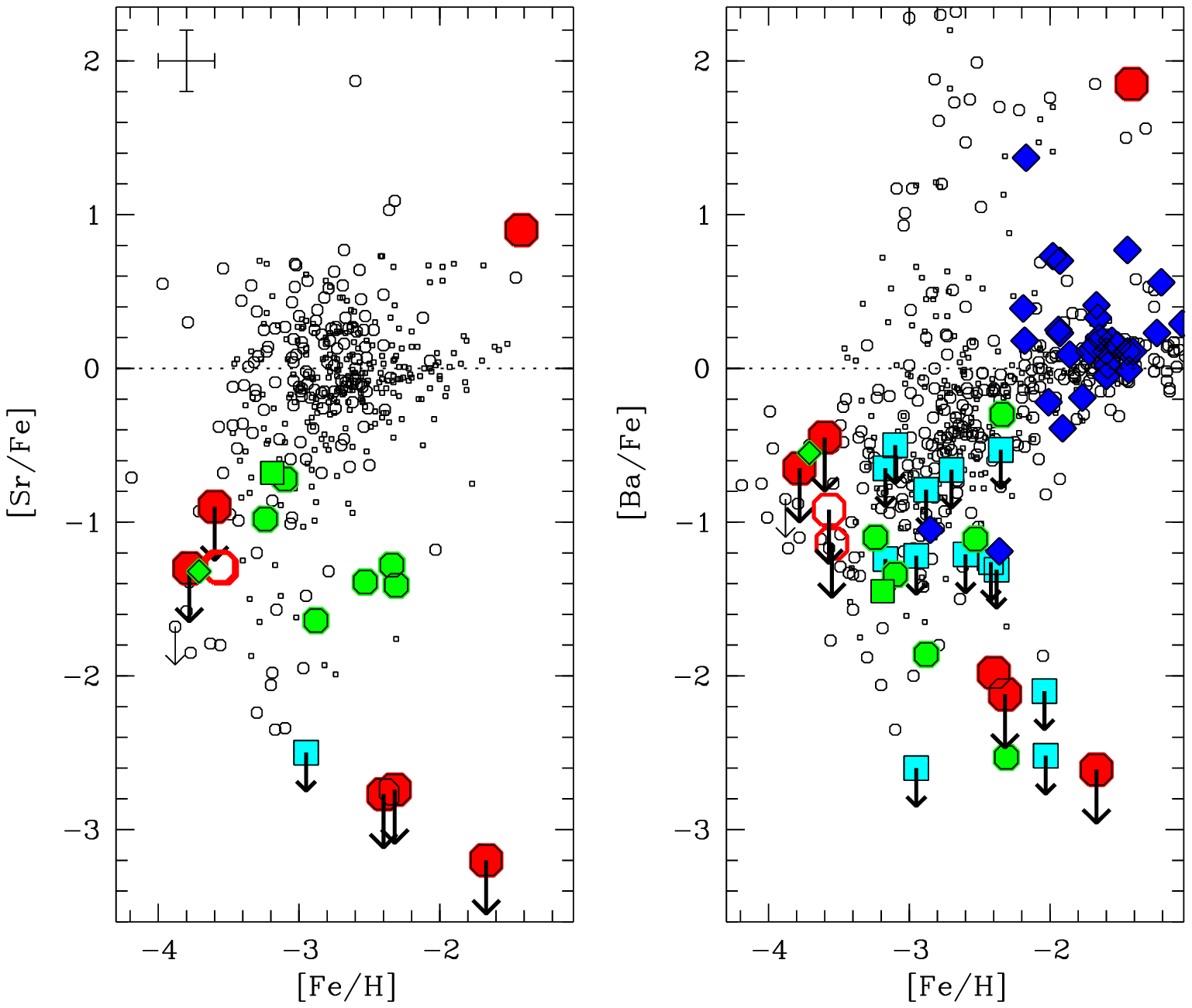}
   \includegraphics[clip=true,width=8.5cm,bbllx=35, bblly=163,
     bburx=385, bbury=508]{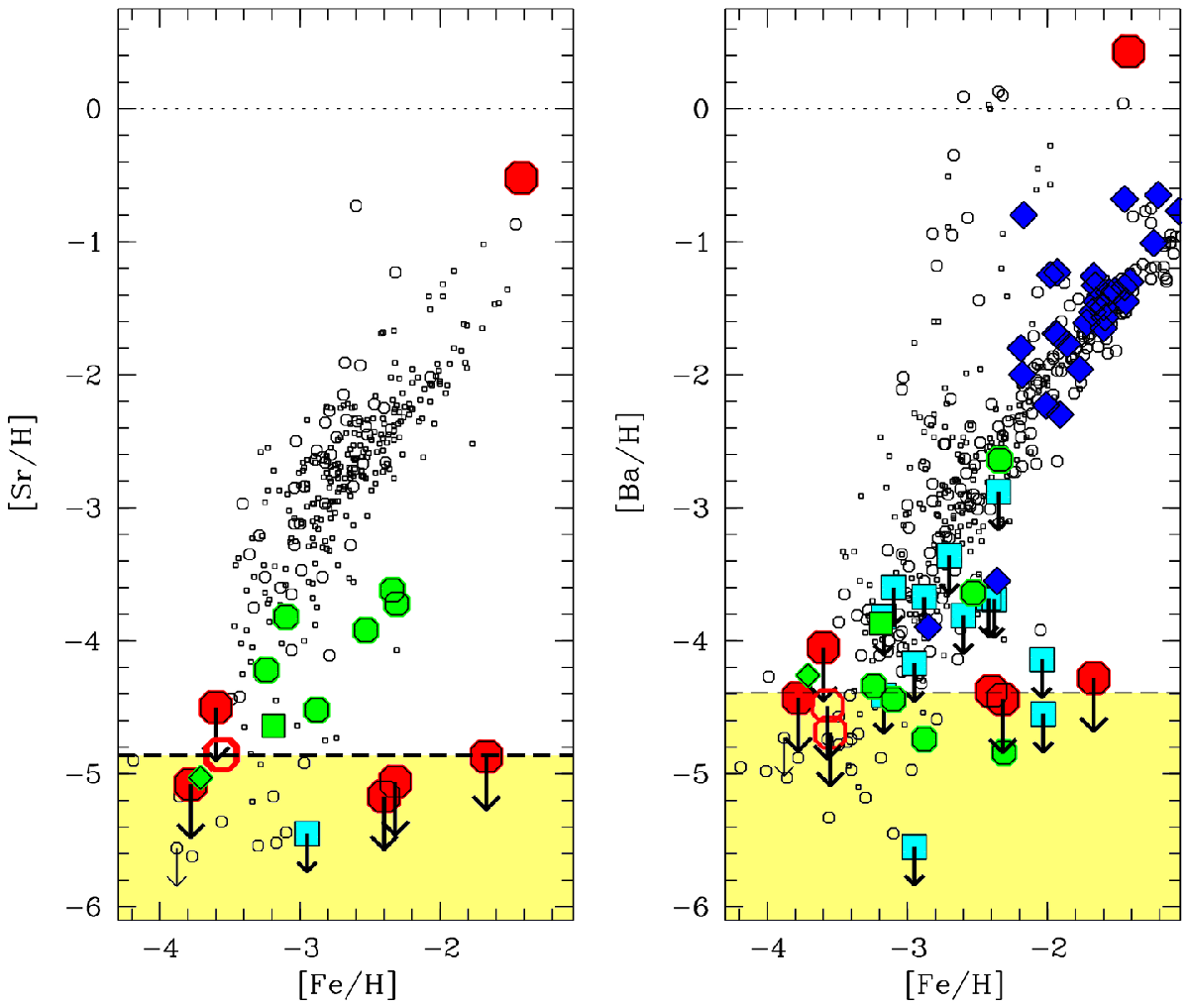}
   \figcaption{\label{ncap_plot} \small Abundance ratios of
     neutron-capture-elements [Sr/Fe] and [Ba/Fe] (top panel) and
     [Sr/H] and [Ba/H] (bottom panel) as a function of metallicity
     [Fe/H] of our Segue\,1 stars (\textit{filled red circles}) and
     star Segue\,1-7 (\textit{open red circle};
     \citealt{norris10_seg}) in comparison with those of other ultra
     faint dwarf galaxy stars in UMa\,II, ComBer, Leo\,IV
     (\textit{green circles}; \citealt{ufs, leo4}), Draco and Hercules
     (\textit{cyan squares}; \citealt{fulbright_rich, koch_her,
       koch13}), stars in the classical dwarfs (\textit{blue
       triangles}; \citealt{venn04}), and the galactic halo
     (\textit{black open circles}; \citealt{aoki05},
     \citealt{francois07}, \citealt{heresII} and
     \citealt{yong13_II}). Note that both axes have the same scale,
     showing the huge range of neutron-capture abundances in
     metal-poor stars. A representative error bars is shown in the
     [Sr/H] panel.}
  \end{center}
 \end{figure}

To complete the discussion on the neutron-capture elements, we also
have to revisit SDSS\,J100714+160154. At $\mbox{[Fe/H]}=-1.4$ it has
$\mbox{[Sr/Fe]}=+0.90$ and $\mbox{[Ba/Fe]}=+1.85$. These high values
are difficult to reconcile with those of the other Segue\,1 stars
unless SDSS\,J100714+160154 has a different enrichment history. We
suggest that SDSS\,J100714+160154 is likely a CH star in a binary
system that underwent mass transfer. General CH star
characteristics are moderate metallicity around
$\mbox{[Fe/H]}\sim-1.5$, high carbon abundance in combination with
s-process enrichment, and radial velocity variation.  The radial
velocity measured from our MIKE spectrum agrees within the
uncertainties with those determined by \citet{simon11}, so binary
amplitudes larger than $\sim10$\,km\,s$^{-1}$ are unlikely.  Despite
the lack of evidence for binary orbital motion, SDSS\,J100714+160154
does have $\mbox{[Fe/H]}\sim-1.4$, $\mbox{[C/Fe]}=1.4$, and also
$\mbox{[Pb/Fe]}\sim2.0$. High lead abundances are a signature of the
s-process, and together with the overabundance in carbon,
SDSS\,J100714+160154's abundance pattern is clearly consistent with
that of a CH star.  In Figure~\ref{seg100_sproc} we show the few
measurable neutron-capture element abundances in SDSS\,J100714+160154
compared to the solar r- and s-process patterns \citep{2000burris}.
While the number of neutron-capture element abundance is limited, the
pattern matches that of the solar s-process.  Hence, the star's
neutron-capture material is not a reflection of the birth gas cloud
composition but is due to a later-time external enrichment event
following mass transfer from the binary companion. While our sample is
too small for a statistical analysis and also lacks long-term radial
velocity monitoring, finding one binary in a sample of seven stars
suggests that the binary fraction is substantial, consistent with the
results of \citet{martinez11}. Binary stars, and thus mass transfer
events, are very common in general, so it is perhaps not too
surprising that a star in our sample shows these characteristics.

\subsection{Abundance uncertainties} \label{unc}

Given that we use our abundances to constrain a number of processes
related to early star formation, chemical enrichment and metal mixing,
it is important to carefully consider any abundance
uncertainties. Using a formula in \citet{o_he1327} to estimate
  the uncertainties on our equivalent width measurements (taking into
  account the $S/N$ of the spectrum, pixel width and number of pixels
  across a spectral line), we estimate our measurements to have
  3$\sigma$ uncertainties of 10 to 15\%. This corresponds to up to 0.15
  to 0.2\,dex (and in the case of SDSS J100742+160106 to up to
  $\sim$0.3\,dex) changes in abundances. These uncertainties are
  reflected in the standard deviations of our line abundances for each
  element (see Table~\ref{Tab:Eqw}) which range from 0.10 to
  0.25\,dex, and are listed in Table~\ref{err}. We also derived
  systematic uncertainties of our abundances for two example stars by
  changing one stellar parameter at a time by an amount approximately
  equal to its random uncertainty. 
  Uncertainties from different sources of random errors (std. dev.
  of the line abundances for a given element), systematic errors,
  and the total uncertainty for each element are given in Table~\ref{err}. 
  Overall, typical total uncertainties range from 0.15 to 0.30\,dex.  

Many abundances are based on only a few lines so that the associated
standard errors are unrealistically small ($<0.05$\,dex) compared with
typical measurement uncertainties. We thus adopt a minimum uncertainty
of 0.10\,dex in such cases. Furthermore, for abundances of elements
derived from only one line, we adopt a formal uncertainty of
0.15\,dex, which should reasonably reflect our largest source of
  error, continuum placement uncertainties. Despite having two
measurements for C, we adopt 0.20\,dex for this element since the
continuum placement was difficult for spectral regions with molecular
bands. Finally, uncertainties on iron abundances are also somewhat
larger. This is due to our method of correcting the spectroscopic
temperatures, which introduces a slope of line abundance as a function
of excitation potential and thus inflates the standard deviation.

\begin{deluxetable}{lrrrrr} 
\tablecolumns{3} 
\tablewidth{0pc} 
\tablecaption{\label{err} Example Abundance Uncertainties for SDSS\,J100710+160623 and SDSS\,J100652+160235}
\tablehead{\colhead{Element}&\colhead{Standard}& 
\colhead{$\Delta$\mbox{T$_{\rm eff}$}}&\colhead{$\Delta\log g$}& 
\colhead{$\Delta v_{micr}$}&\colhead{Total\tablenotemark{a}}\\ 
\colhead{}&\colhead{deviation}&\colhead{+100\,K}& 
\colhead{$+$0.3\,dex}&\colhead{+0.3\,km\,s$^{-1}$}&\colhead{Unc.}\\\hline
\multicolumn{5}{c}{SDSS\,J100710+160623}}
\startdata 
%      ran   teff  logg    vmic     tot                                        
 C (CH)& 0.20 & 0.02 &$-$0.03 &   0.12 & 0.24 \\ 
 Na\,I & 0.10 & 0.14 &$-$0.12 &$-$0.10 & 0.23 \\
 Mg\,I & 0.10 & 0.10 &$-$0.10 &$-$0.08 & 0.19 \\ 
 Al\,I & 0.25 & 0.14 &$-$0.13 &$-$0.11 & 0.33 \\
 Si\,I & 0.15 & 0.07 &$-$0.01 &$-$0.06 & 0.18 \\
 Ca\,I & 0.23 & 0.09 &$-$0.09 &$-$0.12 & 0.29 \\

 Sc\,II& 0.12 & 0.02 &   0.10 &$-$0.09 & 0.18 \\
 Ti\,I & 0.15 & 0.15 &$-$0.04 &$-$0.12 & 0.25 \\
 Ti\,II& 0.15 & 0.02 &   0.09 &$-$0.12 & 0.21 \\
 Cr\,I & 0.16 & 0.13 &$-$0.02 &$-$0.08 & 0.22 \\
 Cr\,II& 0.15&$-$0.04&   0.10 &$-$0.03 & 0.19 \\
 Mn\,I & 0.15 & 0.14 &$-$0.05 &$-$0.13 & 0.25 \\
 Fe\,I & 0.22 & 0.12 &$-$0.04 &$-$0.12 & 0.28 \\
 Fe\,II& 0.18 & 0.03 &   0.10 &$-$0.08 & 0.22 \\
 Co\,II& 0.07 & 0.17 &$-$0.02 &$-$0.18 & 0.26 \\
 Ni\,I & 0.16 & 0.10 &$-$0.01 &$-$0.06 & 0.20 \\\hline
\multicolumn{5}{c}{SDSS\,J100652+160235}\\\hline
 C (CH)& 0.20 & 0.08 &$-$0.05 &   0.12 & 0.25 \\
 Na\,I & 0.10 & 0.10 &$-$0.03 &$-$0.04 & 0.15 \\ 
 Mg\,I & 0.06 & 0.11 &$-$0.09 &$-$0.04 & 0.16 \\ 
 Al\,I & 0.20 & 0.08 &$-$0.01 &$-$0.03 & 0.21 \\ 
 Si\,I & 0.15 & 0.13 &$-$0.11 &$-$0.10 & 0.25 \\ 
 Ca\,I & 0.13 & 0.09 &$-$0.05 &$-$0.05 & 0.15 \\ 
 Sc\,II& 0.15 & 0.08 &   0.08 &$-$0.10 & 0.21 \\ 
 Ti\,II& 0.14 & 0.06 &   0.09 &$-$0.05 & 0.18 \\ 
 Cr\,I & 0.15 & 0.09 &$-$0.01 &$-$0.02 & 0.18 \\ 
 Fe\,I & 0.23 & 0.13 &$-$0.05 &$-$0.12 & 0.30 \\ 
 Co\,II& 0.10 & 0.09 &   0.00 &$-$0.02 & 0.14 \\ 
 Ni\,I & 0.15 & 0.14 &$-$0.05 &$-$0.15 & 0.26 
\enddata 
\tablenotetext{a}{Obtained by adding all uncertainties in quadrature.}
\end{deluxetable}

\section{On the origin and evolution of Segue\,1}\label{history}

We now discuss the main chemical abundance signatures to further
characterize the nature of Segue\,1. Specifically, we aim at testing
whether Segue\,1 is a surviving first galaxy. If so, Segue\,1
  could be a surviving member of a population of the earliest building
  blocks that were available for galaxy formation in the early
  Universe.  We also use Segue\,1 to learn about the origin of the
  most metal-poor stars, assuming that the (outer) Milky Way halo
  assembled from smaller galaxies over time.

\vspace{0.3cm}
\noindent \textbf{Iron abundance spread: consequences of inhomogeneous
  mixing of earliest metals} Segue\,1 contains stars that span a large
range in metallicity of more than 2\,dex, from $\mbox{[Fe/H]}=-1.4$ to
$\mbox{[Fe/H]}=-3.8$. Given the paucity of observable stars as well as
this large Fe spread, the metallicity distribution function -- insofar
as this term can be applied here -- appears to be rather flat. If the
true metallicity distribution is roughly Gaussian
\citep[e.g.,][]{kirby11}, the dispersion must be very large
($\sim1$\,dex).  However, if the distribution reflects inhomogeneous
mixing at early times rather than steady evolution (see below),
then a Gaussian form may not be expected.

According to hydrodynamical simulations by \citet{greif11}, who
studied metal enrichment in a first galaxy, large abundance spreads of
several dex in [X/H] are found already after the explosion of just one
energetic supernova.  The Fe spread in the simulations, averaging
around $\mbox{[Fe/H]}\sim-3$, agrees very well with what we observe in
Segue\,1. Furthermore, one important consequence arises from this: in
an early galaxy that has only undergone an initial chemical enrichment
event but no chemical \textit{evolution} yet, the iron abundance of
the stars does not (yet) provide the kind of time sequence that iron
usually does in other galaxies, including the more luminous classical
dwarf galaxies.  

Ordinarily, as star formation and chemical enrichment proceed in a
system, increasing numbers of stars with progressively higher
metallicities are formed.  These stars then build up the familiar
metallicity distribution function \citep[e.g.,][]{kirby11} with large
numbers of relatively metal-rich stars and a small metal-poor tail
that formed at early times.  Hidden in the metal-rich end of the MDF,
though, should be a few metal-rich stars that were produced at the
earliest times as a consequence of inhomogeneous mixing, but
identifying such stars would be difficult underneath the dominant
younger population at similar metallicities. (Implications of this
scenario are further discussed below in relation to the
neutron-capture element abundances.)  Consequently, the metal-poor
tail of an integrated galaxy MDF consists solely of the few metal-poor
stars that formed after the first enrichment event(s) and before the
system experienced further enrichment.  Rather quickly, the overall
galaxy metallicity becomes too high to form additional metal-poor
stars, and subsequent generations of stars are added only at the
metal-rich end of the distribution.  In Segue\,1, however, this later
enrichment may never have occurred.

This would imply that the most metal-poor Galactic stars are a
collection of second (and/or perhaps third) generation stars that were
born in various small dwarf galaxies. The nature and similarity of
their overall elemental abundance patterns reflects massive star
progenitors, albeit with indications for some variety among the
individual progenitor stars, as would be expected from stochastically
sampling the upper end of the IMF. Given that these small systems
likely formed from a small number of minihalos that hosted Pop\,III
stars (not necessarily with the same mass or mass function), slight
variations in the abundance ratios can be understood in terms of
variations in early Pop\,III stars and the first galaxy assembly
processes.

We note that simulations predict that a fraction of the Milky Way's
stellar halo is likely to have formed in situ, rather than being
accreted from smaller galaxies (e.g., \citealt{zolotov09}).  The
chemical abundance patterns resulting from in situ star formation will
differ from those seen in the very different environments of dwarf
galaxies \citep{brusadin13}, which would alter our expectations for
the comparison between dwarf galaxy stars and halo stars. However,
because in situ stars are thought to be heavily weighted toward the
center of the Milky Way and high metallicities (e.g.,
\citealt{zolotov10}; \citealt{tissera12}), they probably do not 
comprise a significant part of the halo samples to which we 
compare.

\vspace{0.3cm}
\noindent \textbf{Carbon enhancement: massive rotating progenitors
  seeding the first low-mass stars} The three most metal-poor stars in
Segue\,1 are carbon-enhanced.  This presence of significant amounts of
carbon above a critical metallicity \citep{dtrans} likely enabled the
formation of the now-observed low-mass stars. Even the more metal-rich
stars have high enough [C/H] values in agreement with cooling by
  carbon and other metals.  Moreover, the [C/H] ratios of all stars
do not have a clear trend with [Fe/H], suggesting that the production
of these elements and their subsequent mixing is decoupled.  If (some
of) the massive progenitor stars were rapidly rotating, CNO material
would have been released prior to their supernovae, resulting in a
carbon-enhancement floor, and hence no strong correlation with the
supernova-provided elements such as Fe.  Consequently, [C/Fe] shows a
correlation with [Fe/H], with the most metal-poor stars being more
carbon-rich.

\begin{figure}[!th]
 \begin{center}
   \includegraphics[clip=true,width=9.3cm,bbllx=150, bblly=245,
     bburx=450, bbury=620]{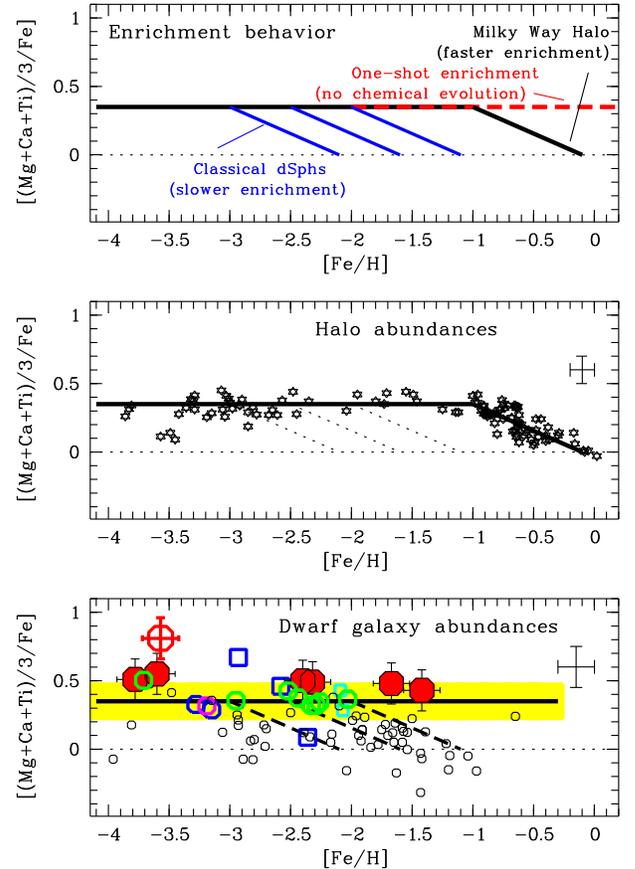}
   \figcaption{ \label{oneshot} Combined Mg-Ca-Ti $\alpha$-element
     abundances as diagnostic of early star formation. adapted from
     \citet{frebel12}.  Top: Schematic representation of chemical
     enrichment in the [$\alpha$/Fe] vs [Fe/H] plane for different
     environments. The dotted line indicates the solar ratio. Middle:
     High-resolution $\alpha$-abundances of metal-poor stars from
     Cayrel et al. (2004) (halo) and Fulbright (2000) (thin/thick
     disk). The diagonal dotted lines indicate the enrichment behavior
     of the dSph galaxies (see top panel). A representative
     uncertainty is shown. Bottom: High-resolution $\alpha$-abundances
     of metal-poor stars in the classical dSphs (\textit{small open
       black circles} and several evolutionary paths are indicated with
     dashed lines; see \citealt{frebel12} and references therein).
     Different colors denote different ultra-faint dwarf
     galaxies. \textit{Filled red circles}: Segue\,1 stars from this
     study; \textit{open red circle}: Segue\,1-7 
     \citep{norris10_seg}; \textit{blue squares}: Coma Berenices;
     \textit{blue circles}: Ursa Major II; \textit{pink circle}: Leo
     IV; \textit{cyan squares}: Hercules; \textit{green circles}:
     Bo\"otes I.  The yellow shaded region around
     $\mbox{[$\alpha$/Fe]} = 0.35$ depicts the predicted
     \citet{frebel12} one-shot enrichment behavior (with a 0.15\,dex
     observational uncertainty) reflecting massive core-collapse
     supernova enrichment.}
 \end{center}
\end{figure}

\vspace{0.3cm}
\noindent \textbf{Enhanced $\alpha$-element abundances: no late-time
  star formation}
The $\alpha$-element abundance ratios indicate what type of stars
provided the observed elements.  The top panel of Figure~\ref{oneshot}
schematically depicts the differences between the chemical evolution
and enrichment history of various galaxies: the Milky Way halo, the classical
dwarfs with slower chemical evolution, and the early systems with
essentially no evolution at all.  Segue\,1 stars all have enhanced
$\mbox{[$\alpha$/Fe]}$ values consistent with chemical enrichment of
their birth gas cloud by massive stars, with no contribution from
other sources of heavy elements. We show the combined
$\mbox{[$\alpha$/Fe]}$ abundances for the Segue\,1 stars in the bottom
panel of Figure~\ref{oneshot} (red circles). On the other hand, the
Milky Way halo (middle panel) was enriched rapidly to
$\mbox{[Fe/H]}\approx-1$ by core-collapse supernovae before Type\,Ia
supernovae began to add significant amounts of iron.  In the classical
dSphs (lower panel, black circles), the Type\,Ia contribution (which
is assumed to occur after a similar delay time in all galaxies) is
visible at lower metallicities, $-2.5 \lesssim \mbox{[Fe/H]} \lesssim
-1.5$ \citep{venn04,letarte10,kirby11b,lemasle12}, indicating that
less enrichment took place during the pre-Ia epoch.

The fact that Segue\,1 shows enhanced [$\alpha$/Fe] abundances for all
metallicities reflects rapid and inhomogeneous metal mixing,
consistent with the predictions of \citet{frebel12} for the chemical
signature of a first or very early galaxy: the observed stars should show
no sign of AGB star or supernova Ia enrichment indicating star
formation beyond the second generation. The general outline of
  the evolution of such a system is as follows.  First, Pop\,III stars
  form (perhaps only one per minihalo) and nearly instantly enrich
  their surroundings with some metals.  Next, a second generation of
stars formed that included long-lived low-mass stars. The higher mass
stars of this generation soon exploded as core collapse supernovae,
while the intermediate-mass stars went through their AGB phase to
release newly created elements through stellar winds. These supernova
explosions were likely powerful enough to blow out gas from  the
  shallow potential wells of these early systems, preventing further
star formation. Alternatively, star formation could also have been
suppressed soon after the formation of the galaxy by reionization
(e.g., \citealt{lunnan11}), or both.

%======================================================

\vspace{0.3cm}
\noindent \textbf{Low neutron-capture element abundances: evidence for
  one progenitor generation?}
All Segue\,1 stars that have not been contaminated by a binary
companion display an extremely low content of neutron-capture
material. Considering stars with $\mbox{[Fe/H]}<-3$ first, their
[Sr/H] and [Ba/H] ratio upper limits and the Sr abundance measurement
of Segue\,1-7 equal those of the lowest ones observed in Galactic halo
stars. These low abundances indicate a total amount of each
neutron-capture element of no more than $\sim10^{-7}$\,M$_{\odot}$ in
Segue\,1. A single neutron-capture element producing event such as an
r-process supernova or a massive rotating Pop\,III star undergoing
s-process nucleosynthesis may produce $10^{-4}$\,M$_{\odot}$ of
neutron-capture elements \citep{farouqi10,pignatari}, suggesting that
at most one such event occurred during the star forming epoch in
Segue\,1. Hence, specific progenitor properties such as a particular
mass range of supernova undergoing the r-process or perhaps an early
black-hole forming supernova with little to no neutron-capture element
yields could have led to such a low overall level of neutron-capture
elements in Segue\,1.  Deeper observations of Segue\,1 stars that
  measure neutron-capture abundances rather than providing upper
  limits offer the possibility of constraining the yield of this
  event.

If Segue\,1 is a surviving first galaxy, then no intermediate-mass AGB
star enrichment signatures should be discernible in its stellar
abundances. In the halo, s-process material provided by AGB stars only
occurs at $\mbox{[Fe/H]}>-2.7$ \citep{simmerer2004}, suggesting a
corresponding characteristic neutron-capture element abundance level
of $\mbox{[Sr/H]}>-3$ and $\mbox{[Ba/H]}>-3.5$.  The Sr and Ba
abundances in Segue\,1 appear to be $\sim2$\,dex lower than that level
even for the more metal-rich stars, confirming that AGB stars did not
contribute to the enrichment of the gas cloud from which the Segue\,1
stars formed.

Based on the similar high abundances of $\alpha$-elements and low
abundances of neutron-capture elements, it is plausible that a
significant fraction of the most metal-poor halo stars ($\mbox{[Fe/H]}
\lesssim -3$) were formed in first galaxies like Segue\,1.  Estimating
solely from the overlap between the halo [Sr/H] abundances and those
in the ultra-faint dwarfs, up to half of EMP halo stars could have
originated in systems with the same degree of neutron-capture element
depletion.  The halo stars with $\mbox{[Sr/H]} \gtrsim -4$ and
$\mbox{[Ba/H]} \gtrsim -4$ likely formed in galaxies with larger
stellar masses that underwent more chemical evolution with
contributions from AGB stars.  This picture is in qualitative
agreement with that proposed by \citet{lee_d_13} to explain the
difference in Ba and Sr abundance distributions between the
ultra-faint dwarfs and the halo.

Interestingly, the Sr and Ba abundances of the higher metallicity
Segue\,1 stars are just as low as those of the extremely metal-poor
stars.  There are no halo stars known at $\mbox{[Fe/H]} \gtrsim -2.5$
that have similarly low [Sr/H] and [Ba/H], which means that the
fraction of halo stars in this metallicity range that originated in
first galaxies must be negligibly small.  However, this does not mean
that first galaxies cannot have contributed to the buildup of the halo
at all.  As described above in the Fe spread discussion, the fraction
of (relatively) metal-rich stars now found in a larger galaxy like the
Milky Way that formed in a first galaxy is expected to be minuscule.
Their small numbers should be vastly overshadowed by the huge
population of later-generation stars with similar metallicities formed
in the many more-evolved dwarf galaxies that were eventually
incorporated into the Galactic halo.  A prediction of this scenario is
that large enough surveys of the Milky Way (e.g., GALAH, Gaia) should
identify a few fossil second-generation metal-rich stars with
extremely low neutron-capture abundances.  Assuming that the Segue\,1
abundance pattern is representative of the class of first galaxies,
halo stars with $\mbox{[Fe/H]} > -2.5$, $\mbox{[$\alpha$/H]} \ge 0.4$,
$\mbox{[Ba/H]} \le -4$, and $\mbox{[Sr/H]} \le -5$ should be
considered as candidates for having originated in a first galaxy.  If
such stars can be found, they would provide additional evidence for
inhomogeneous mixing of metals in the earliest galaxies and the
hierarchical assembly of the Galaxy from these kinds of systems.

Finally, keeping the effects of inhomogeneous metal mixing in mind,
abundance spreads [X/H] should be present in all elements, not just
Fe. Since we only have 3$\sigma$ upper limits for [Sr/H] and [Ba/H]
with the exception of one Sr and one Ba detection (in two different
stars), we cannot make firm statements about the spread in
neutron-capture element abundances in Segue\,1. However, if our
postulation above regarding the origin of the most neutron-capture
depleted halo stars is correct, the existence of Milky Way stars
at $\mbox{[Sr/H]}\sim-5.6$, which is $\sim0.7$\,dex below the lone Sr
detection in Segue\,1, suggests that there could be a detectable
spread in neutron-capture abundances in Segue\,1. The presence
  of a large spread in Sr and Ba is also indicated when combining the
  Segue\,1 results with those of the most metal-poor stars in other
  ultra-faints such as Coma Berenices, Ursa Major\,II and Leo\,IV
  \citep{ufs,leo4}, as shown in Figure~\ref{ncap_plot}.  However, only
  much higher S/N spectra of new and existing stars (which will be
  extremely difficult, if not impossible, to obtain with current
  telescopes) will allow us to conclusively test this prediction.

\vspace{0.3cm} 
\noindent \textbf{Segue\,1 as an ancient surviving first galaxy}
Considering the detailed chemical abundances of the seven brightest
stars in Segue\,1 to describe the origin and evolution of this galaxy
thus suggests that no significant chemical \textit{evolution}, and
hence star formation, has taken place in Segue\,1 since its formation.
There is no indication of AGB star or supernova Ia enrichment prior to
the birth of any of the observed stars. It thus appears that only the
first star yields have been preserved in the atmospheres of the
now-observed long-lived stars. The resulting abundance pattern of
similarly metal-poor halo stars closely resembles that of our Segue\,1
stars, suggesting that the metal-poor tail of the Galactic metallicity
distribution function might have been assembled by numerous early
dwarf galaxies that produced the most metal-poor stars in their
respective second or early generations of stars before additional
supernovae added further metals.

The surviving first galaxy model \citep{frebel12} predicts an ancient
single-age stellar population, i.e., the long-lived part of the second
generation of stars of an early galaxy that formed from minihalos in
the early Universe. In terms of their chemical composition, the
ultra-faint dwarfs so far all show very similar characteristics
(although this conclusion is currently based on only one or a few
observable stars in each galaxy). Indeed, besides Segue\,1, Leo\,IV,
Coma Berenices, and Bo\"otes\,I are on the candidate first galaxy list
of \citet{frebel12}. Assuming Segue\,1 and other ultra-faint dwarf
galaxies to be surviving first galaxies then also suggests that they
would need to be very old.  \citet{brown12} and \citet{brown13}
recently presented age measurements based on deep HST color-magnitude
diagrams of three ultra-faint dwarf galaxies (Hercules, Leo\,IV and
Ursa Major\,I) and preliminary results for Coma Berenices, Bo\"otes\,I
and Canes Venatici\,II. They find all of them to show very similar,
single-age populations that are at least as old as the globular
cluster M92 with $\sim13\pm$1\,Gyr.

Hence, the Brown et al. studies confirm the predicted old age of the
ultra-faint dwarfs, including at least three galaxies proposed by
\citet{frebel12} as surviving first galaxies and a majority of those
suggested by \citet{bovill09} as fossils (although using a different
definition). While Segue\,1 contains too few stars for a robust star
formation history to be derived from deep CMDs, the apparently
identical stellar populations of the other ultra-faint dwarfs and the
consistency of available data with the possibility of a single-age
population in Segue\,1 constitute strong clues that Segue\,1 is
similarly old and thus shows all expected signatures of a surviving
first galaxy.

\vspace{0.3cm}
\noindent \textbf{The History of Supernova Explosions in Segue\,1} A
key determinant of the early evolution of Segue\,1 is how many
supernova explosions it hosted during its star-forming epoch.  The
number of supernovae, in turn, depends on the IMF.  Segue\,1 does not
contain enough stars for reliable IMF measurements from deep
\emph{HST} photometry \citep[e.g.,][]{brown12,geha13}, but one can
nevertheless attempt an estimate.  \citet{martin08} determined that
Segue\,1 contains $65 \pm 9$ stars brighter than $r = 22$.  According
to a 13~Gyr, $\alpha$-enhanced, $\mbox{[Fe/H]} = -2.49$ isochrone from
\citet{dotter08}, this magnitude limit corresponds to a stellar mass
of 0.70~M$_{\odot}$, and the main sequence turnoff occurs at
0.79~M$_{\odot}$.  Using these numbers to set the normalization for
the IMF, if the IMF above 0.5~M$_{\odot}$ has the \citet{salpeter}
slope of $\alpha=2.35$ and at lower masses follows the shallower slope
($\alpha=1.3$) determined by \citet{kroupa93}, then the present-day
stellar mass of Segue\,1 from the hydrogen-burning limit up to the
turnoff is 700\,M$_{\odot}$, and the initial stellar mass before all
of the massive stars evolved was 1500\,M$_{\odot}$.  When using such a
bottom-heavy IMF, the upper mass cutoff assumed makes hardly any
difference; a 50\,M$_{\odot}$ vs. 100\,M$_{\odot}$ cutoff changes the
initial stellar mass by less than 4\%.  The total number of stars more
massive than 8\,M$_{\odot}$ that would be expected to explode as
supernovae is $\sim15$, again with little dependence on the upper mass
cutoff of the IMF.

However, \citet{geha13} have shown that the ultra-faint dwarfs
Hercules and Leo\,IV have a significantly bottom-light IMF for
subsolar mass stars.  It is not necessarily the case that the shallow
power-law slope they measure for low-mass stars continued unbroken to
$M>8$\,M$_{\odot}$ at early times, but in the absence of evidence to
the contrary, it is interesting to explore the consequences of such an
assumption.  With the \citeauthor{geha13} IMF, the stellar mass of
Segue\,1 today is 500\,M$_{\odot}$, and the initial stellar mass was
$\sim10^{4}$\,M$_{\odot}$.  Unlike the Kroupa case, the upper mass
cut of this very top-heavy IMF has a large impact on the initial mass.
Increasing the maximum mass from 50 to 100\,M$_{\odot}$ changes the
initial mass by 50\%.  The number of massive stars with this top-heavy
IMF is very large: $\sim300$ (250-400 for upper mass cutoffs between
30 and 100\,M$_{\odot}$).

It is then interesting to investigate what the potential metal yield
of these supernovae might have been. Taking the average metal mass
ejected by one core-collapse supernova to be $\sim3$\,M$_{\odot}$
yields 900\,M$_{\odot}$ of metals synthesized by supernovae in
Segue\,1. for the Geha IMF, and 45\,M$_{\odot}$ of metals for the
Kroupa IMF.  Because the observed heavy element content of the stars
in the galaxy today is $\sim0.01$\,M$_{\odot}$, in either case the
vast majority (99.98 to 99.999\%) of metals produced by Segue\,1
supernovae must have been blown out of the galaxy.  Such extremely
efficient outflow of metals is consistent with (although more extreme
than) the trends seen in more luminous dwarf galaxies by
\citet{kirby11c}.  In order to avoid incorporating these metals into
the low-mass stars in Segue\,1, this may suggest that the duration of
star formation was shorter than the lifetimes of most of the massive
stars ($<40$\,Myr), and perhaps that star formation in Segue\,1 was
shut off by one of the first few supernovae that occurred.  

Another way of testing whether these IMF scenarios are realistic is to
consider the energy deposit from supernovae to assess whether the
system would be disrupted entirely.  Following \citet{whalen08} and
\citet{johnson13}, $<10^{7}$\,M$_{\odot}$ halos are destroyed by one
massive (pair-instability) supernova with an explosion energy of
$10^{53}$\,ergs, and $10^{7}$\,M$_{\odot}$ halos by
$10^{54}$\,ergs. The destruction of a more massive halo, such as a
$10^{8}$\,M$_{\odot}$ atomic cooling halo (a halo that can be
considered a first galaxy), would require more than $10^{54}$\,ergs.

For a typical supernova explosion of $10^{51}$\,ergs, at least
$\sim1000$ supernovae would be required to disrupt even a
$10^{7}$~M$_{\odot}$ halo, suggesting that Segue\,1 could have
survived the supernovae associated with a top-heavy IMF.  This
conclusion is strengthened by the fact that the abundance patterns of
the most metal-poor halo stars favor ``faint'', lower energy supernova
explosions \citep{UmedaNomotoNature, iwamoto_science}, of $3 \times
10^{50}$\,ergs.  Only if most of the explosions were substantially
higher energy hypernovae could $\sim300$ massive stars potentially
destroy the galaxy.  Moreover, we have assumed that the coupling of
the supernova explosion energy to gas kinetic energy is 100\%
efficient.  The coupling efficiency depends on the exact configuration
of the gas, but a more reasonable order of magnitude is 10\%
\citep{thornton98}.  Again, this would imply that hundreds of
supernovae could have been present in Segue\,1 at early times.
However, even though the galaxy's potential well was deep enough to
survive a large number of supernovae, these explosions would have had
dramatic effects on its gas content.  

 We therefore do not find any inconsistency with the hypothesis that
 the early-time Segue\,1 could have had a top-heavy IMF, similar to
 what has been found for Hercules and Leo\,IV \citep{geha13}.
 Ultimately, more data as well as modeling of such early galaxies will
 hopefully reveal the nature of the IMF in these surviving ancient
 systems.

\section{Prospects and Conclusion}\label{sec:conc}

We have presented chemical abundance measurements of six stars in the
faint Segue\,1 dwarf galaxy. Together with Segue\,1-7
\citep{norris10_seg}, these are the brightest cool stars --- and
the known red giants --- in Segue\,1, having magnitudes around
$g\sim19$\,mag. Thus, they are the only ones for which high-resolution
spectra of sufficient quality can be obtained with current
telescopes. Given that this sample is brightness-selected, the
metallicity spread of nearly 2.5\,dex is remarkable. Overall,
metallicities range from $\mbox{[Fe/H]}=-1.4$ to $\mbox{[Fe/H]}=-3.8$,
and three of the seven stars have $\mbox{[Fe/H]}\lesssim-3.5$,
indicating that Segue\,1 contains the highest fraction of extremely
metal-poor stars ever seen.

The chemical abundances show that Segue\,1 stars over the full
observed metallicity range have enhanced $\alpha$-element abundances
at the level of metal-poor halo stars, indicating enrichment only from
massive stars.   Segue\,1 is the first (and only) galaxy in which
  no decline in [$\alpha$/Fe] is seen at higher metallicities.  The
three most metal-poor stars are enhanced in carbon, also pointing to
massive progenitors. The extremely small amounts of neutron-capture
elements found in all Segue\,1 stars point to a single neutron-capture
material production event in association with massive stars, rather
than AGB stars. Altogether, the abundance signature of Segue\,1 agrees
with predictions for Segue\,1 being a surviving first galaxy that
underwent only one generation of star formation after its formation
from $\sim10$ Pop\,III star hosting minihalos \citep{frebel12}.

If Segue\,1 is indeed a surviving fossil, then its properties can help
us understand the origin and nature of the most metal-poor stars in
the Galactic halo. Specifically, simulations of the first galaxies
suggest that these small, early systems could only produce extremely
metal-poor stars as part of their second generation, i.e. the first
generation after formation in which low-mass stars could be made. At
later times, if any gas remains, it will no longer be metal-poor due
to fast mixing of the metals ejected by the early supernovae. An
implication of this scenario would be that many of the most
metal-poor stars in the Galaxy stem from their respective first/early
galaxies and are in all likelihood second generation stars. This
would, to some extent, explain why it has been so difficult to model
the low-metallicity tail of the halo metallicity distribution function
with chemical evolution models (e.g., \citealt{schoerck}).

Following the abundance trends found in Segue\,1, it appears that high
[$\alpha$/Fe] abundances together with extremely low neutron-capture
abundances may be a tell-tale sign of those first galaxy
second-generation stars. Many of the most metal-poor halo stars indeed
share this signature with the Segue\,1 stars especially at the
  lowest [Fe/H] values. Since early inhomogeneous mixing leads to
large spreads in elements [X/H], the large spread in neutron-capture
element abundances in the Milky Way may to some extent
describe the same inhomogeneity that we observe in Segue\,1 in terms
of its [Fe/H] spread. But instead of reflecting the content of just
one galaxy, it is the superposition of spreads from the many galaxies
that contributed to the build up of the metal-poor stellar population
of the Galaxy.

If inhomogeneous mixing is responsible for the observed abundance
spreads, the same level of scatter in neutron-capture abundance ratios
should be observed in the most metal-poor stars in the classical
dwarfs. Moreover, these systems should contain a relatively higher
fraction of those early second generation stars than the halo. Indeed,
Hercules (with $5\times10^{4}\,L_{\odot}$) and Draco (with
$4\times10^{5}\,L_{\odot}$), are known to contain stars with unusually
low neutron-capture abundances. One star in Draco at
$\mbox{[Fe/H]}=-2.9$ has upper limits of
$\mbox{[Sr/H]}\sim\rm{[Ba/H]}<-5.5$, measured from a relatively high
$S/N$ spectrum by \citep{fulbright_rich}.It also has elevated, near
halo-like [$\alpha$/Fe] values ($\mbox{[Mg/Fe]}=0.50$,
$\mbox{[Ca/Fe]}=-0.07$, $\mbox{[Ti/Fe]}=0.21$).  The two stars in
Hercules at $\mbox{[Fe/H]}=-2.1$ have
$\mbox{[Sr/H]}\sim{\rm{[Ba/H]}}<-4.2$ and similar [$\alpha$/Fe]
abundances ($\mbox{[Mg/Fe]}=0.79$, $\mbox{[Ca/Fe]}=-0.11$,
$\mbox{[Ti/Fe]}=0.17$ for Her-2, and $\mbox{[Mg/Fe]}=0.75$,
$\mbox{[Ca/Fe]}=0.21$, $\mbox{[Ti/Fe]}=0.33$ for Her-3
\citep{koch_her}.

Interestingly, no satisfactory explanations for the origin of the low
neutron-capture element abundance stars in Draco and Hercules have
been found until now \citep{fulbright_rich, koch_her}. But if one were to
consider these extreme stars as ``left over stars'' from their host
system's own building blocks one could possibly explain their
existence by assuming that systems with $\sim10^{5}\,L_{\odot}$ and
more are already ``assembled galaxies'' themselves.  If systems like
Draco show true second-generation stars (here taken to have very low
neutron-capture element abundances) from their respective first galaxy
building blocks at a rate of only 1 in 15 (based on the observed
sample of stars in Draco; \citealt{shetrone01, fulbright_rich, cohen09}),
then it could be understood why no such stars have yet been found in
the even more luminous classical dwarf galaxies, and especially the
halo.

These dwarf spheroidal galaxies also show declining
$\alpha$-abundances with increasing metallicity \citep{kirby11b}.
These higher metallicity stars must be from later stellar generations
which presumably  after mergers with additional of their
building blocks or after significant gas accretion and
retention. Broadly speaking, the lower $\alpha$-abundance stars
reflect later-time enrichment by supernova Ia (e.g.,
\citealt{shetrone01}) although other scenarios may be able to explain
the observations of the stellar content of these dwarfs as
well. Detailed analyses of more stars at all metallicities, but
particularly at high metallicity in the ultra-faint dwarfs and at
low-metallicity in the classical dwarfs, will reveal more about the
assembly histories of dwarf galaxies as well as the formation of the
halo of the Milky Way.

%\newpage

\begin{deluxetable}{lrrrrrrrrrrrrrrrrrrrrrrrrrrrr}
\tabletypesize{\tiny}
\tablewidth{0pc}
\tablecaption{\label{abund} Magellan/MIKE Chemical Abundances of all Segue\,1 Stars}
\tablehead{
\colhead{Species} & 
\colhead{$N$} &
\colhead{$\log\epsilon (\mbox{X})$} & \colhead{$\sigma$}&  \colhead{[X/H]}& \colhead{[X/Fe]}} 
\startdata
\multicolumn{5}{c}{SDSS\,J100714+160154}\\\hline\\
%# Element   Num.  log(X)    sigma(X)  [X/H]     [X/Fe]
  CH    &   1 &    8.45 &    0.20 &    0.02 &    1.44 \\
  Na\,I &   2 &    4.79 &    0.12 & $-$1.46 & $-$0.04 \\
  Mg\,I &   7 &    6.56 &    0.18 & $-$1.04 &    0.38  \\
  Al\,I &   2 &    4.37 &    0.20 & $-$2.08 & $-$0.66 \\
  Si\,I &   1 &    6.48 &    0.15 & $-$1.03 &    0.39 \\
  Ca\,I &  22 &    5.36 &    0.13 & $-$0.98 &    0.44 \\
 Sc\,II &   2 &    1.68 &    0.17 & $-$1.47 & $-$0.05 \\
  Ti\,I &  25 &    3.84 &    0.11 & $-$1.11 &    0.31 \\
 Ti\,II &  28 &    4.01 &    0.10 & $-$0.94 &    0.48 \\
  Cr\,I &  16 &    4.21 &    0.19 & $-$1.43 & $-$0.02 \\
 Cr\,II &   2 &    4.68 &    0.10 & $-$0.96 &    0.46 \\
  Mn\,I &   5 &    3.51 &    0.21 & $-$1.92 & $-$0.50 \\
  Fe\,I & 187 &    6.08 &    0.24 & $-$1.42 &    0.00 \\
 Fe\,II &  23 &    6.08 &    0.18 & $-$1.42 & $-$0.00 \\
  Co\,I &   4 &    3.65 &    0.17 & $-$1.34 &    0.08 \\
  Ni\,I &  10 &    4.79 &    0.12 & $-$1.43 & $-$0.01 \\
  Zn\,I &   2 &    3.29 &    0.22 & $-$1.27 &    0.15  \\
 Sr\,II &   1 &    2.35 &    0.20 & $-$0.52 &    0.90 \\
 Zr\,II &   1 &    2.56 &    0.25 & $-$0.02 &    1.40 \\
 Ba\,II &   3 &    2.61 &    0.18 &    0.43 &    1.85 \\
 La\,II &   1 &    1.48 &    0.25 &    0.38 &    1.80 \\
 Eu\,II &   3 & $-$0.17 &    0.14 & $-$0.69 &    0.73 \\
 Pb\,I  &   1 &    2.30 &    0.50 &    0.55 &    1.97 \\\hline 
\multicolumn{5}{c}{SDSS\,J100710+160623}\\\hline 
  CH    &   2 &    6.91 &    0.20 & $-$1.52 &    0.15 \\ 
  Na\,I &   2 &    4.31 &    0.10 & $-$1.93 & $-$0.26 \\
  Mg\,I &   5 &    6.39 &    0.10 & $-$1.21 &    0.47 \\
  Al\,I &   1 &    3.88 &    0.25 & $-$2.57 & $-$0.90 \\ 
  Si\,I &   1 &    6.37 &    0.15 & $-$1.14 &    0.53 \\
  Ca\,I &  25 &    5.25 &    0.23 & $-$1.09 &    0.58 \\
 Sc\,II &   7 &    1.61 &    0.12 & $-$1.54 &    0.13 \\
  Ti\,I &  26 &    3.66 &    0.15 & $-$1.29 &    0.38 \\
 Ti\,II &  39 &    3.65 &    0.15 & $-$1.30 &    0.38 \\
  Cr\,I &  17 &    3.84 &    0.16 & $-$1.80 & $-$0.13 \\
 Cr\,II &   1 &    3.76 &    0.15 & $-$1.88 & $-$0.21 \\
  Mn\,I &   6 &    3.27 &    0.15 & $-$2.16 & $-$0.49 \\
  Fe\,I & 210 &    5.83 &    0.22 & $-$1.67 &    0.00 \\
 Fe\,II &  23 &    5.85 &    0.18 & $-$1.65 &    0.02 \\
  Co\,I &   5 &    3.32 &    0.07 & $-$1.67 &    0.00 \\
  Ni\,I &  14 &    4.56 &    0.16 & $-$1.66 &    0.01 \\
  Zn\,I &   1 & $<$2.84 & \nodata & $<-$1.72&$<-$0.05 \\
 Sr\,II &   1 & $<-$2.00& \nodata & $<-$4.87& $<-$3.20\\
 Ba\,II &   1 & $<-$2.10& \nodata & $<-$4.28& $<-$2.61\\
 Eu\,II &   1 & $<-$1.15& \nodata & $<-$1.67&  $<$0.00\\\hline
\multicolumn{5}{c}{SDSS\,J100702+155055}\\\hline 
  CH    &   2 &    6.19 &    0.20 & $-$2.24 &    0.08 \\   %alpha/fe=0.55+0.42+0.57+0.43=0.49
  Na\,I &   2 &    4.07 &    0.14 & $-$2.17 &    0.15 \\
  Mg\,I &   8 &    5.80 &    0.12 & $-$1.80 &    0.52 \\
  Al\,I &   2 &    3.29 &    0.20 & $-$3.16 & $-$0.84 \\
  Si\,I &   2 &    5.60 &    0.15 & $-$1.90 &    0.42 \\
  Ca\,I &  19 &    4.53 &    0.11 & $-$1.81 &    0.51 \\
 Sc\,II &   5 &    0.93 &    0.15 & $-$2.22 &    0.10 \\
  Ti\,I &  11 &    3.01 &    0.11 & $-$1.94 &    0.38 \\
 Ti\,II &  35 &    3.05 &    0.17 & $-$1.90 &    0.43 \\
  Cr\,I &  10 &    3.17 &    0.09 & $-$2.47 & $-$0.15 \\
  Mn\,I &   7 &    2.54 &    0.15 & $-$2.87 & $-$0.57 \\
  Fe\,I & 162 &    5.18 &    0.18 & $-$2.32 &    0.00 \\ 
 Fe\,II &  15 &    5.18 &    0.15 & $-$2.32 &    0.00 \\ 
  Co\,I &   2 &    2.57 &    0.10 & $-$2.42 & $-$0.10 \\
  Ni\,I &   5 &    3.79 &    0.23 & $-$2.43 & $-$0.11 \\
  Zn\,I &   1 & $<$2.69 & \nodata &$<-$1.87 &  $<$0.45\\
 Sr\,II &   1 & $<-$2.19& \nodata &$<-$5.06 & $<-$2.74\\
 Ba\,II &   1 & $<-$2.26& \nodata &$<-$4.44 & $<-$2.12\\
 Eu\,II &   1 & $<-$1.50& \nodata &$<-$2.02 &  $<$0.30\\\hline
\enddata \tablenotetext{a}{[X/H] and [X/Fe] values have been
  recalculated from $\log$ gf values using the \citet{asplund09} solar
  abundances.}
\end{deluxetable}

\begin{deluxetable}{lrrrrrrrrrrrrrrrrrrrrrrrrrrrr}
\tabletypesize{\tiny}
\tablewidth{0pc}
\tablecaption{\label{abund} Table 5 continued -- Magellan/MIKE Chemical Abundances of all Segue\,1 Stars}
\tablehead{
\colhead{Species} &
\colhead{$N$} &
\colhead{$\log\epsilon (\mbox{X})$} & \colhead{$\sigma$}&  \colhead{[X/H]}& \colhead{[X/Fe]}}
\startdata
\multicolumn{5}{c}{SDSS\,J100742+160106}\\\hline         
% Element Num.  log(X)    sigma(X)  [X/H]     [X/Fe]  
  CH    &   2 &    5.93 &    0.20 & $-$2.50 & $-$0.10 \\
  Na\,I &   2 &    3.61 &    0.10 & $-$2.63 & $-$0.23 \\
  Mg\,I &   7 &    5.74 &    0.08 & $-$1.86 &    0.54 \\
  Al\,I &   2 &    3.38 &    0.15 & $-$3.07 & $-$0.67 \\
  Si\,I &   1 &    5.74 &    0.15 & $-$1.77 &    0.63 \\
  Ca\,I &  21 &    4.46 &    0.09 & $-$1.88 &    0.52 \\
 Sc\,II &   6 &    1.06 &    0.10 & $-$2.09 &    0.31 \\
  Ti\,I &  16 &    2.95 &    0.10 & $-$2.00 &    0.40 \\
 Ti\,II &  33 &    3.00 &    0.13 & $-$1.95 &    0.45 \\
  Cr\,I &   9 &    2.96 &    0.17 & $-$2.68 & $-$0.28 \\
  Mn\,I &   4 &    2.61 &    0.25 & $-$2.82 & $-$0.42 \\
  Fe\,I & 170 &    5.10 &    0.17 & $-$2.40 &    0.00 \\
 Fe\,II &  17 &    5.11 &    0.10 & $-$2.39 &    0.01 \\
  Co\,I &   2 &    2.72 &    0.14 & $-$2.27 &    0.13 \\
  Ni\,I &   1 &    3.95 &    0.15 & $-$2.27 &    0.13 \\
  Zn\,I &   1 & $<$2.47 & \nodata &$<-$2.09 & $<$0.31  \\ 
  Sr\,II&   1 &$<-$2.30 & \nodata &$<-$5.17 & $<-$2.77 \\
  Ba\,II&   1 & $-$2.20 &    0.20 & $-$4.38 &  $-$1.98 \\
  Eu\,II&   1 &$<-$1.40 & \nodata &$<-$1.92 & $<$0.48  \\  
\multicolumn{5}{c}{SDSS\,J100652+160235}\\\hline 
% Element Num.  log(X)    sigma(X)  [X/H]     [X/Fe]  
  CH    &   2 &    6.03 &    0.20 & $-$2.40 &    1.20 \\  
  Na\,I &   2 &    2.75 &    0.10 & $-$3.49 &    0.11 \\
  Mg\,I &   6 &    4.59 &    0.06 & $-$3.01 &    0.59 \\
  Al\,I &   1 &    2.25 &    0.20 & $-$4.20 & $-$0.60 \\
  Si\,I &   1 &    4.31 &    0.15 & $-$3.19 &    0.41 \\
  Ca\,I &   3 &    3.30 &    0.13 & $-$3.04 &    0.56 \\
 Sc\,II &   1 & $-$0.10 &    0.15 & $-$3.25 &    0.35 \\ 
 Ti\,II &  12 &    1.95 &    0.14 & $-$3.00 &    0.60 \\
  Cr\,I &   1 &    1.72 &    0.15 & $-$3.92 & $-$0.32 \\
  Mn\,I &   1 & $<$1.50 & \nodata &$<-$3.93 &$<-$0.33 \\
  Fe\,I &  47 &    3.90 &    0.23 & $-$3.60 &    0.00 \\
  Co\,I &   2 &    1.91 &    0.10 & $-$3.08 &    0.52 \\
  Ni\,I &   4 &    2.75 &    0.15 & $-$3.47 &    0.13 \\
  Zn\,I &   1 & $<$2.40 & \nodata & $<-$2.16& $<$1.44 \\ 
 Sr\,II &   1 & $<-$1.63& \nodata & $<-$4.50& $<-$0.90\\
 Ba\,II &   1 & $<-$1.87& \nodata & $<-$4.05& $<-$0.45\\
 Eu\,II &   1 & $<-$1.08& \nodata & $<-$1.60&  $<$2.00\\\hline
\multicolumn{5}{c}{SDSS\,J100639+160008}\\\hline 
% Element  Num.  log(X)    sigma(X)  [X/H]     [X/Fe] 
  CH    &   2 &    5.56 &    0.25 & $-$2.87 &    0.91 \\  %alpha/fe=0.63+0.40+0.51+0.54=0.52
  Na\,I &   2 &    2.74 &    0.10 & $-$3.50 &    0.28 \\
  Mg\,I &   4 &    4.40 &    0.10 & $-$3.20 &    0.57 \\
  Al\,I &   2 &    1.98 &    0.25 & $-$4.47 & $-$0.70 \\
  Si\,I &   1 &    5.00 &    0.25 & $-$2.51 &    1.27 \\
  Ca\,I &   2 &    3.16 &    0.10 & $-$3.18 &    0.59 \\
 Sc\,II &   2 & $-$0.65 &    0.10 & $-$3.80 & $-$0.02 \\
 Ti\,II &  12 &    1.55 &    0.18 & $-$3.40 &    0.37 \\
  Cr\,I &   3 &    1.44 &    0.16 & $-$4.20 & $-$0.42 \\
  Mn\,I &   2 &    1.10 &    0.10 & $-$4.33 & $-$0.55 \\
  Fe\,I &  35 &    3.72 &    0.11 & $-$3.78 &    0.00 \\
  Co\,I &   2 &    1.61 &    0.10 & $-$3.38 &    0.40 \\
  Ni\,I &   3 &    2.59 &    0.12 & $-$3.63 &    0.15 \\
  Zn\,I &   1 & $<$2.04 & \nodata &$<-$2.52 & $<$1.26 \\
 Sr\,II &   1 & $<-$2.21& \nodata &$<-$5.08 &$<-$1.30 \\ 
 Ba\,II &   1 & $<-$2.25& \nodata &$<-$4.43 &$<-$0.65 \\
 Eu\,II &   1 & $<-$1.26& \nodata &$<-$1.78 & $<$2.00 \\\hline
\multicolumn{5}{c}{Segue\,1-7 (from Norris et al. 2010a)\tablenotemark{a}}\\\hline
% Element  Num.  log(X)    sigma(X)  [X/H]     [X/Fe]                
  CH    & 2 &    7.17& 0.20  & $-$1.26 &    2.31 \\  
  Na\,I & 2 &    3.18& 0.04  & $-$3.06 &    0.51 \\
  Mg\,I & 5 &    4.95& 0.07  & $-$2.65 &    0.92 \\
  Al\,I & 1 &    3.08&\nodata& $-$3.37 &    0.20 \\
  Si\,I & 1 &    4.79& 0.20  & $-$2.72 &    0.85 \\
  Ca\,I & 9 &    3.63& 0.04  & $-$2.71 &    0.86 \\
 Sc\,II &\nodata&\nodata&    & \nodata & \nodata \\
 Ti\,II & 11&    2.03& 0.08  & $-$2.92 &    0.65 \\
  Cr\,I & 3 &    1.86& 0.05  & $-$3.78 & $-$0.21 \\
  Mn\,I & 1 &    1.31&\nodata& $-$4.12 & $-$0.55 \\
  Fe\,I & 37&    3.93& 0.03  & $-$3.57 &    0.00 \\
  Co\,I & 4 &    1.77& 0.15  & $-$3.22 &    0.35 \\
  Ni\,I & 1 &    2.16&\nodata& $-$4.06 & $-$0.49 \\
  Zn\,I &\nodata&\nodata&    & \nodata & \nodata \\
 Sr\,II & 2 & $-$1.99& 0.23  & $-$4.86 & $-$1.29 \\
 Ba\,II & 1 &$<-$2.31&\nodata&$<-$4.49 &$<-$0.92 \\
 Eu\,II & 1 &$<-$2.20&\nodata&$<-$2.72 & $<$0.85 \\\hline
\enddata \tablenotetext{a}{[X/H] and [X/Fe] values have been
  recalculated from $\log$ gf values using the \citet{asplund09} solar
  abundances.}
\end{deluxetable}

\acknowledgements{We thank Andrew McWilliam for providing the Arcturus
  spectrum. A.F. is supported by NSF CAREER grant
  AST-1255160. J.D.S. is supported by NSF grant AST-1108811.
  E.N.K. acknowledges support from the Southern California Center for
  Galaxy Evolution, a multicampus research program funded by the
  University of California Office of Research. This work made use of
  the NASA's Astrophysics Data System Bibliographic Services.

We are grateful to the many people who have worked to make the Keck
Telescope and its instruments a reality and to operate and maintain
the Keck Observatory. The and to operate and maintain the Keck
Observatory. The authors wish to extend special thanks to those of
Hawaiian ancestry on whose sacred mountain we are privileged to be
guests. Without their generous hospitality, none of the
observations presented herein would have been possible.  }

\textit{Facilities:} \facility{Magellan-Clay (MIKE), Keck:I (HIRES)}

\end{document}

%% file: equivalent_widths_segue_stars.tex
\begin{deluxetable*}{lllrrrrrrrrrrrrr}
%\rotate                                                                                                        
\tablecolumns{8}
\tablewidth{0pt}
\tabletypesize{\footnotesize}
\tabletypesize{\tiny}
\tablecaption{\label{Tab:Eqw} Equivalent width measurements of the Segue\,1 stars  (Table will be available electronically)}
\tablehead{
\colhead{El.} & \colhead{$\lambda$} &\colhead{$\chi$} &\colhead{$\log gf$}&
\colhead{EW [m{\AA}]}&\colhead{$\lg\epsilon$ [dex]}&
\colhead{EW [m{\AA}]}&\colhead{$\lg\epsilon$ [dex]}&
\colhead{EW [m{\AA}]}&\colhead{$\lg\epsilon$ [dex]}&
\colhead{EW [m{\AA}]}&\colhead{$\lg\epsilon$ [dex]}&
\colhead{EW [m{\AA}]}&\colhead{$\lg\epsilon$ [dex]}&
\colhead{EW [m{\AA}]}&\colhead{$\lg\epsilon$ [dex]}\\
\colhead{}&\colhead{[{\AA}]}&\colhead{[eV]}&\colhead{[dex]}&
\multicolumn{2}{c}{SDSS\,J100714+160154}&
\multicolumn{2}{c}{SDSS\,J100710+160623}&
\multicolumn{2}{c}{SDSS\,J100702+155055}&
\multicolumn{2}{c}{SDSS\,J100742+160106}&
\multicolumn{2}{c}{SDSS\,J100652+160235}&
\multicolumn{2}{c}{SDSS\,J100639+160008}   \\
}
\startdata
%        &  {\AA}    & eV   &         &  m{\AA} &   dex   & m{\AA}    &   dex   & m{\AA}  &  dex    & m{\AA}  &   dex   &  m{\AA} &  dex    &  m{\AA} &  dex \\
  CH     &4313     &\nodata&\nodata & syn    & \nodata &   syn     &   6.86  &   syn   &   6.19  &   syn   &    5.83 &  syn    & 5.98    &   syn   &  5.45   \\
  CH     &4325     &\nodata&\nodata & syn    &    8.45 &   syn     &   6.96  &   syn   &   6.19  &   syn   &    6.03 &  syn    & 6.08    &   syn   &  5.67   \\
  Na\,I  &5889.950 & 0.00 &   0.108 & 240.13 &    4.67 &    190.82 &    4.26 &  148.77 &    3.86 &  118.78 &    3.59 &   59.86 &    2.71 &   78.14 &    2.79 \\
  Na\,I  &5895.924 & 0.00 &$-$0.194 & 228.99 &    4.90 &    171.68 &    4.36 &  146.43 &    4.13 &  104.38 &    3.63 &   49.00 &    2.80 &   57.83 &    2.69 \\
  Mg\,I  &3829.355 & 2.71 &$-$0.208 &\nodata & \nodata &   \nodata & \nodata & \nodata & \nodata & \nodata & \nodata &  105.97 &    4.49 & \nodata & \nodata \\
  Mg\,I  &3832.304 & 2.71 &   0.270 &\nodata & \nodata &   \nodata & \nodata & \nodata & \nodata & \nodata & \nodata &  149.84 &    4.53 &  136.37 &    4.37 \\
  Mg\,I  &3986.753 & 4.35 &$-$1.030 & 136.44 &    6.79 &   \nodata & \nodata &   41.40 &    5.57 &   45.98 &    5.71 & \nodata & \nodata & \nodata & \nodata \\
  Mg\,I  &4057.505 & 4.35 &$-$0.890 & 115.68 &    6.52 &    102.39 &    6.39 &   52.37 &    5.61 &   58.80 &    5.79 & \nodata & \nodata & \nodata & \nodata \\
\enddata
\tablecomments{The table is available in its entirety only in
  electronic format. A portion is shown for guidance regarding its
  form and content.}
\end{deluxetable*}